\documentclass[
aip,jcp,reprint]{revtex4-2}
\usepackage[english]{babel}

\usepackage{graphicx}
\usepackage{dcolumn}
\usepackage{bm}

\usepackage[utf8]{inputenc}
\usepackage[T1]{fontenc}
\usepackage{etoolbox}
\usepackage{xcolor}

\usepackage{ragged2e}
\usepackage{amsmath}
\usepackage{amssymb}
\usepackage{booktabs}
\usepackage{multirow}
\usepackage{graphicx}
\usepackage{subcaption}

\usepackage[left = 1.37cm, right = 1.37cm, top = 2cm, bottom = 2cm]{geometry}

\emergencystretch=\maxdimen
\allowdisplaybreaks

\makeatletter
\def\@email#1#2{
 \endgroup
 \patchcmd{\titleblock@produce}
  {\frontmatter@RRAPformat}
  {\frontmatter@RRAPformat{\produce@RRAP{*#1\href{mailto:#2}{#2}}}\frontmatter@RRAPformat}
  {}{}
}
\makeatother
\begin{document}

\preprint{AIP/123-QED}

\definecolor{AppleGreen}{rgb}{0.55, 0.71, 0.0}
\definecolor{OliveGreen}{rgb}{0.1, 0.4, 0.1}
\definecolor{CarmineRed}{rgb}{1.0, 0.0, 0.22}

\newcommand{\AH}[1]{{\it\textcolor{AppleGreen}{#1}}}
\newcommand{\AI}[1]{{\textcolor{OliveGreen}{#1}}}
\newcommand{\ET}[1]{{\it\textcolor{CarmineRed}{#1}}}

\title{Surfactant reorientation under shear: dynamic surface tension and droplet deformation}
\author{Alexandra J.\ Hardy}
 \affiliation{School of Mathematics and Statistics, The Open University, Walton Hall, Milton Keynes MK7 6AA, United Kingdom}
 
\author{Abdallah Daddi-Moussa-Ider}
\affiliation{School of Mathematics and Statistics, The Open University, Walton Hall, Milton Keynes MK7 6AA, United Kingdom}

\author{Elsen Tjhung}
\email{elsen.tjhung@open.ac.uk}
\affiliation{School of Mathematics and Statistics, The Open University, Walton Hall, Milton Keynes MK7 6AA, United Kingdom}

\date{\today}

\begin{abstract}
Surfactants are amphiphilic molecules that are generally anisotropic rather than spherical. 
Their orientation is governed by the interplay between shear-induced rotation, thermal rotational diffusion, and energetic alignment with the interface. 
The relative importance of these processes is characterized by the rotational P\'eclet number, $\text{Pe}_r$.
We show that this microscopic coupling between flow and surfactant orientation can give rise to new macroscopic interfacial phenomena, including a shear-dependent effective surface tension and non-trivial droplet deformation.
To investigate this mechanism, we develop a phase-field model that incorporates both the surfactant concentration and its local average orientation (polarization field). 
Using perturbation theory, we derive an analytical expression for the effective surface tension, which depends not only on the surfactant concentration but also on the local shear rate.
We then employ a hybrid numerical method to study the deformation of a surfactant-covered droplet under imposed shear flow.
For small $\text{Pe}_r$, droplet deformation can be accurately captured by a modified Taylor and Maffettone--Minale theories.
For large $\text{Pe}_r$, shear-induced reorientation strongly distorts the surfactant polarization, and the droplet deformation progressively approaches that of a pure (surfactant-free) droplet. 
\end{abstract}

\maketitle

\section{Introduction}

Surfactants are well known for their ability to lower the surface tension between two immiscible fluids~\cite{mulqueen2002theoretical,mobius2001surfactants}. 
Because of this property, they are widely used across many industrial applications, including detergents, emulsification, food processing, pharmaceuticals, and enhanced oil recovery~\cite{tadros2006applied,schramm2000surfactants,shaban2020surfactants,singh2004potential,nitschke2018recent}. 
In particular, the addition of surfactants gives rise to a wide range of interfacial phenomena~\cite{scamehorn1986overview,rosen2012surfactants,dave2017concise}, including enhanced droplet deformation under an imposed shear flow.

Surfactants consist of a hydrophilic head and a hydrophobic tail and they preferentially align perpendicular to a fluid--fluid interface. 
Under shear flow, however, this preferred orientation can be disrupted, giving rise to a coupling between the local flow and the local surfactant orientation (also called polarization).
At larger scales, this microscopic coupling can dynamically modify macroscopic interfacial properties, including the effective surface tension and droplet deformation.
Using a phase-field model that explicitly incorporates both surfactant concentration and orientation~\cite{hardy2024hybrid,hardy2026kinetic}, 
we demonstrate that an imposed tangential shear tilts the surfactant polarization away from the interface normal. 
This shear-induced reorientation reduces the ability of surfactants to lower the interfacial free energy.
Consequently, although increasing the surfactant concentration decreases the surface tension under equilibrium conditions, increasing the shear rate can dynamically increase the effective surface tension towards its bare value.

This dynamic modification of the surface tension can directly influence droplet deformation. 
For a pure droplet without surfactants, deformation under shear is governed by the competition between viscous and interfacial stresses, quantified by the capillary number, $\text{Ca}$. 
At sufficiently small $\text{Ca}$, Taylor's classical theory predicts that the deformation increases linearly with $\text{Ca}$~\cite{taylor1934formation}. 
At larger $\text{Ca}$, phenomenological models such as the Maffettone--Minale (MM) model provide a nonlinear description of the droplet shape~\cite{maffettone1998equation,minale2010models}. 
Numerical and experimental studies have further identified dynamical phenomena including deformation overshoot, oscillatory relaxation, and eventual breakup at large enough $\text{Ca}$~\cite{sibillo2006drop,feigl2003numerical,cristini2003drop,amani2019numerical,guido2011shear,ioannou2016droplet}.

In the presence of surfactants, the droplet dynamics become more complex due to surfactant redistribution and the resulting Marangoni stresses\cite{janssen1997influence,phillips1980experimental}. 
Previous studies generally report enhanced deformation due to reduced surface tension\cite{stone1990effects,kruijt2004droplet,feigl2007simulation}, with additional effects arising from non-uniform surfactant distributions\cite{milliken1994influence,eggleton1998adsorption,PhysRevFluids.8.090501}. 
However, most existing models treat surfactants as scalar concentration fields and neglect their molecular orientation. 
As a result, they do not capture the coupling between flow-induced surfactant reorientation, interfacial mechanics, and droplet deformation.

To address this limitation, we employ a phase-field model that explicitly incorporates the surfactant orientation and its coupling to both the fluid--fluid interface and the local flow~\cite{hardy2024hybrid,hardy2026kinetic}. 
We first consider a flat interface and show that shear-induced surfactant reorientation dynamically modifies the effective surface tension.
Although increasing the surfactant concentration is known to reduce the effective surface tension, an imposed shear flow weakens this reduction and drives the surface tension back towards its bare value. 
The extent of this effect is controlled by the rotational Péclet number, $\text{Pe}_r$.
 
We then investigate the steady-state deformation of a surfactant-covered droplet under shear flow. 
For small $\text{Pe}_r$, the surfactants remain approximately normal to the interface, allowing a simple analytical expression for the surfactant-renormalized surface tension to be obtained.
In this regime, droplet deformation can be accurately captured by a modified Taylor theory at small $\text{Ca}$ and a modified MM theory at larger $\text{Ca}$.
At larger $\text{Pe}_r$, the local flow significantly distorts the surfactant orientation, and no comparably simple expression for the effective surface tension is available. 
Nevertheless, the simulations show that shear-induced reorientation progressively weakens the surfactant-induced reduction in surface tension, causing the steady-state deformation to approach that of a pure droplet.

Overall, this work establishes a direct link between microscopic surfactant orientational dynamics, the macroscopic effective surface tension, and droplet deformation under shear. 
Our results demonstrate that surfactant orientation constitutes an important dynamical degree of freedom that is absent from conventional scalar descriptions and provide a framework for modelling interfaces coated with anisotropic amphiphilic or colloidal particles under flow.

\section{Phase Field Model\label{sec:model}}

We consider a two-phase fluid system (e.g., oil and water) containing surfactant molecules. 
Surfactants have a hydrophilic head and a hydrophobic tail, causing them to preferentially adsorb at the interface between two fluid phases, typically aligning perpendicular to the interface.
Within a continuum phase-field description, the two fluid phases are described by a single scalar order parameter $\phi(\mathbf{r},t)$, which takes values approximately $+1$ in the oil phase and $-1$ in the water phase. 
The interface between the two fluid phases is diffuse, so that $\phi$ varies smoothly between these two values across a finite interfacial region of thickness $\xi$.
The surfactant molecules are represented by two additional fields: the local surfactant concentration $c(\mathbf{r},t)$ and the average molecular orientation 
$\mathbf{p}(\mathbf{r},t)$, referred to as the polarization field. 
Finally, the binary fluid and surfactant molecules are advected by a common velocity field $\mathbf{u}(\mathbf{r},t)$, 
which satisfies the incompressible Stokes equation.

The thermodynamic behaviour of the system is described by a free-energy functional $\mathcal{F}[\phi,c,\mathbf{p}]$, which captures the coupling between the two fluid phases, the surfactant concentration, and the surfactant orientation\cite{hardy2026kinetic}:
\begin{align}
 \mathcal{F}[\phi,c,\mathbf{p}] &= \int \bigg\{ \frac{\beta}{4}(1-\phi^2)^2 + \frac{\kappa}{2} \, |\nabla\phi|^2 + k_\mathrm{B}T c\ln(\xi^dc) \nonumber \\ 
 &\quad+ \frac{d}{2} \, k_\mathrm{B}T c|\mathbf{p}|^2 + \chi\ell c \mathbf{p}\cdot \nabla\phi \bigg\} \, d^dr \, .
 \label{eq:free-energy}
\end{align}
The first term is the bulk free-energy density, which has minima at $\phi=\pm1$, representing the oil and water phases, with $\beta>0$ setting the energy scale.
The second term represents the interfacial energy, where $\kappa>0$ and $\beta$ together determine the surface tension and the thickness of the diffuse interface.
The third term, proportional to the thermal energy scale $k_\mathrm{B} T$, represents the translational entropy of the surfactant molecules.
The fourth term, also proportional to $k_\mathrm{B} T$, describes the orientational entropy of the surfactants, which favours an isotropic state with $\mathbf{p}\simeq\boldsymbol{0}$.
The final term couples the surfactant orientation to the fluid interface: $\chi>0$ sets the strength of the surfactant-interface interaction, while $\ell>0$ is the characteristic molecular length scale of the surfactants.
Finally, $d$ denotes the spatial dimension of the system. 
In this work, we focus primarily on $d=2$, since the results in $d=3$ are qualitatively similar.

We should note the free energy is quadratic in $|\mathbf{p}|$. 
This is a weak-polarization approximation that permits $|\mathbf{p}|>1$.
In other words, we assume weak coupling $\varepsilon=\chi\ell/k_BT\ll1$.
For the simulations below, we use $\varepsilon=1$ since this does not change the physics.

In the absence of surfactants ($c=0$), the system relaxes toward an equilibrium state that minimizes the free energy in Eq.~\eqref{eq:free-energy}.
For example, one equilibrium configuration is a flat interface located at $y=0$, for which the order parameter is given by $\phi(\mathbf{r})=\tanh(y/\xi)$, where $\xi=\sqrt{2\kappa/\beta}$ defines the interfacial thickness.
Physically, $\xi$ is on the order of molecular length scales, although in numerical simulations it is typically chosen several orders of magnitude larger to ensure stability. From the equilibrium $\phi(\mathbf{r})$, the surface tension in the absence of surfactants is given by $\sigma_0 = \sqrt{8\kappa\beta/9}$. In the presence of surfactants, however, the effective surface tension is modified and can vary locally and dynamically depending on the shear rate (see Section~\ref{sec:tilt}).

The full system dynamics are then described by\cite{hardy2026kinetic}
\begin{subequations} \label{eq:phi-c-p-u}
\begin{align}
  \frac{\partial \phi}{\partial t} + \boldsymbol{\nabla} \cdot(\phi\mathbf{u}) &= M\nabla^2\frac{\delta\mathcal{F}}{\delta\phi} \, ,\label{eq:phidot} \\
  \frac{\partial c}{\partial t} + \boldsymbol{\nabla}\cdot(c\mathbf{u}) &= \boldsymbol{\nabla}\cdot\left(\frac{c}{\Gamma_t}\boldsymbol{\nabla}\frac{\delta\mathcal{F}}{\delta c}\right) \, ,\label{eq:cdot} \\
  \frac{\partial \mathbf{p}}{\partial t} + (\mathbf{u}\cdot \boldsymbol{\nabla})\mathbf{p} &= -\underline{\underline{\boldsymbol{\Omega}}}\cdot\mathbf{p} + \frac{Bd}{d+2}\underline{\underline{\mathbf{D}}}\cdot\mathbf{p} - \frac{d-1}{d\Gamma_r c}\frac{\delta\mathcal{F}}{\delta\mathbf{p}} \, , \label{eq:pdot} \\
0 &= -\boldsymbol{\nabla} P + \eta\nabla^2\mathbf{u} + \mathbf{f} \, , \label{eq:Stokes} \\
0 &= \boldsymbol{\nabla}\cdot\mathbf{u} \, \label{eq:incompressibility},  
\end{align}
\end{subequations}
where $\eta$ is the fluid viscosity (assumed to be equal for the two fluid phases).
We have also defined the strain rate tensor $\underline{\underline{\mathbf{D}}}$ and the vorticity tensor $\underline{\underline{\boldsymbol{\Omega}}}$ as:
\begin{subequations}
\begin{align}
D_{\alpha\beta} &= \frac{1}{2}(\partial_\alpha u_\beta + \partial_\beta u_\alpha) \, , \\
\Omega_{\alpha\beta} &= \frac{1}{2}(\partial_\alpha u_\beta - \partial_\beta u_\alpha) \, .
\end{align}
\end{subequations}
In Eqs.~(\ref{eq:phidot}-\ref{eq:pdot}), $M>0$ sets the mobility of the binary fluid, while $\Gamma_t>0$ and $\Gamma_r>0$ denote the translational and rotational friction coefficients of the surfactant molecules, respectively (assumed to be isotropic and equal for the two fluid phases).
These friction coefficients are related to the corresponding translational and rotational diffusion coefficients through the Einstein relations
\begin{equation}
D_t = \frac{k_BT}{\Gamma_t} \quad\text{and}\quad D_r = \frac{k_BT}{\Gamma_r},
\end{equation}
respectively.
The parameter $B\in[0,1]$ in Eq.~(\ref{eq:pdot}) is the Bretherton constant, which characterizes the rotation of non-spherical objects in a fluid flow.
If surfactants are approximated as ellipsoids with aspect ratio~$\Delta$, the Bretherton constant is given by $B=(\Delta^2-1)/(\Delta^2+1)$.
Thus, $B=0$ corresponds to spherical particles, while $B=1$ corresponds to infinitely thin needle-like particles.
Finally, $\mathbf{f}=\mathbf{g}+\boldsymbol{\nabla}\cdot\underline{\underline{\boldsymbol{\sigma}}}'$ in Eq.~(\ref{eq:Stokes}) is the hydrodynamic force density which depends on $\phi$, $c$, and $\mathbf{p}$: 
\begin{subequations} \label{eq:hydrodynamic-force}
\begin{align}
g_\alpha &= -\phi\partial_\alpha\frac{\delta\mathcal{F}}{\delta\phi} - c\partial_\alpha\frac{\delta\mathcal{F}}{\delta c}
		- p_\beta \partial_\alpha \frac{\delta\mathcal{F}}{\delta p _\beta} \, , \\
\sigma_{\alpha\beta}' &= \frac{Bd}{2(d+2)} \left(p_\alpha\frac{\delta\mathcal{F}}{\delta p_\beta} + p_\beta\frac{\delta\mathcal{F}}{\delta p_\alpha}\right)  \nonumber\\
  &\quad - \frac{1}{2} \left(p_\alpha\frac{\delta\mathcal{F}}{\delta p_\beta} - p_\beta\frac{\delta\mathcal{F}}{\delta p_\alpha}\right) \, .
\end{align}
\end{subequations}
Eqs.~(\ref{eq:phi-c-p-u}) are solved numerically in a two-dimensional domain $(x,y)\in[0,L_x]\times[0,L_y]$. 
Periodic boundary conditions are applied in the $x$-direction, 
while impermeability and free-slip boundary conditions are enforced at the walls: $y=0$ and $y=L_y$. 
The external shear flow is then superimposed, as described in Sec.~\ref{sec:spectral}.

We define the average surfactant concentration as
\begin{equation}
c_0 = \frac{1}{V_d}\int c(\mathbf{r},t)\,d^dr \, ,
\end{equation}
where $V_d$ is the $d$-dimensional volume of the system.
$c_0$ is constant due to the continuity equation (\ref{eq:cdot}).

It is worth noting that, aside from the free energy structure in Eq.~(\ref{eq:free-energy}), the dynamics of $\phi$, $c$, $\mathbf{p}$, and $\mathbf{u}$ in Eqs.~(\ref{eq:phi-c-p-u}) and (\ref{eq:hydrodynamic-force}) are largely identical to those of polar liquid crystal emulsions~\cite{Bonelli2019,negro2019,Cates_Tjhung_2018}.
In related models~\cite{kay2026orientablesurfactantsliquidfilms,hohenberg1977theory}, thermal noise can also be made explicit by introducing Gaussian white noise in the dynamics (\ref{eq:phi-c-p-u}).

\section{Numerical Scheme\label{sec:numerics}}

We take the bare interfacial thickness $\xi=\sqrt{2\kappa/\beta}$ as the unit of length and the thermal energy $k_\mathrm{B}T$ as the unit of energy. 
In these units, $\kappa=\beta/2$, $\Gamma_r=D_r^{-1}$, and $k_\mathrm{B}T=1$. 
We also define a small dimensionless parameter $\varepsilon=\chi\ell/(\xi k_\mathrm{B}T)$ to represent the weak coupling between the fluid and surfactant. In these dimensionless units, the free energy (\ref{eq:free-energy}) becomes~\cite{hardy2026kinetic}:
\begin{align}
 \mathcal{F}[\phi,c,\mathbf{p}] &= \int \bigg\{ \frac{\beta}{4}(1-\phi^2)^2 + \frac{\beta}{4}\, |\nabla\phi|^2 + c\ln(c) \nonumber \\ 
 &\quad+ \frac{d}{2} \, c|\mathbf{p}|^2 + \varepsilon c \mathbf{p}\cdot \nabla\phi \bigg\} \, d^dr \, .
 \label{eq:free-energy-dimensionless}
\end{align}
The dynamics remain governed by Eqs.~(\ref{eq:phi-c-p-u}) with $\Gamma_r=D_r^{-1}$ now represents the rotational diffusion timescale of the surfactant molecules.

Numerically, the spatial coordinates $\mathbf{r}=(x,y)$ are discretized as $x=i\Delta x$ and $y=j\Delta x$, where $i=0,1,2,\dots,N_x-1$ and $j=0,1,2,\dots,N_y-1$.
Here, $\Delta x$ and $\Delta y$ denote the spatial grid spacings along the $x$- and $y$-directions, respectively.
The physical lengths of the domain are then $L_x = N_x \Delta x$ and $L_y = N_y \Delta y$. 
On the other hand, time is discretized into $t = 0, \Delta t, 2\Delta t, \dots$, where $\Delta t$ is the time step.
In all simulations we fix $\Delta x = \Delta y = 0.5$ and $\Delta t = 0.001$. 
The fields $\phi$, $c$, and $\mathbf{p}$ are evolved using a finite-difference scheme with forward Euler time stepping. 
The velocity field $\mathbf{u}$ is computed using a spectral method. 
All numerical computations are implemented in Python using the NumPy\cite{harris2020array} and SciPy\cite{virtanen2020scipy} libraries.

\subsection{Finite Difference Method}

Equations~(\ref{eq:phidot}-\ref{eq:pdot}) are solved using a finite-difference scheme with forward Euler time stepping. Spatial derivatives are approximated using a finite-difference stencil~\cite{pooley2008eliminating}. For a generic field value $q^{i,j}$, where $q$ may represent $\phi$, $c$, $p_x$, etc., the first-order spatial derivatives are computed as follows:
\begin{subequations}
\begin{align}
  \frac{\partial q^{i,j}}{\partial x} &\simeq \frac{1}{12 \Delta x}\left(q^{i+1,j+1} - q^{i-1,j+1} + 4q^{i+1,j} \right.\nonumber\\ &\quad \left.- 4q^{i-1,j} + q^{i+1,j-1} - q^{i-1,j-1}\right),\label{eq:numeric-dqdx}\\
  \frac{\partial q^{i,j}}{\partial y} &\simeq \frac{1}{12 \Delta y}\left(q^{i+1,j+1} - q^{i+1,j-1} + 4q^{i,j+1} \right.\nonumber\\ &\quad\left.- 4q^{i,j-1} + q^{i-1,j+1} - q^{i-1,j-1}\right).\label{eq:numeric-dqdy}
\end{align}
\end{subequations}
It should be noted that $\Delta x = \Delta y$ in practice. 
The Laplacian is computed as follows:
\begin{align}
  \nabla^2q^{i,j} &\simeq \frac{1}{6\Delta x\Delta y} \Big(q^{i-1,j+1} + 4q^{i,j+1} + q^{i+1,j+1} \nonumber \\ 
  &\quad+ 4q^{i-1,j}  -20q^{i,j} + 4q^{i+1,j}  \nonumber \\ 
  &\quad + q^{i-1,j-1}+ 4q^{i,j-1} + q^{i+1,j-1} \Big) \, . \label{eq:numeric-lap}
\end{align}
Here, we use a 9-point stencil improve isotropy in the discrete gradient operator.

We apply neutral wetting boundary conditions for $\phi$ and $c$ at the walls $y=0$ and  $y=L_y$. 
First, we introduce a set of ghost points at $j=-1$ and $j=N_y$.
The boundary conditions for $q=\phi$, $c$,  at these ghost points are then given by:
\begin{equation}
q^{i,j = -1} = q^{i,j = 0} \quad\text{and}\quad q^{i,j = N_y} = q^{i,j = N_y - 1} \, , \label{eq:no-flux-1}
\end{equation}
for all $i$.
We apply the same boundary conditions for $p_x$ and $p_y$ (free anchoring conditions).

Since Eqs.~(\ref{eq:phidot}) and (\ref{eq:cdot}) can be written in the form of continuity equations $\partial_t\phi+\boldsymbol{\nabla\cdot\mathbf{J}}_\phi=0$ and $\partial_tc+\boldsymbol{\nabla\cdot\mathbf{J}}_c=0$, 
we also need to impose no-flux/reflecting boundary conditions on the currents.
Specifically, for a generic current component $q=J_{\phi,x}$, $J_{\phi,y}$, $J_{c,x}$ or $J_{c,y}$ we apply:
\begin{equation}
q^{i,j = -1} = -q^{i,j = 0} \quad\text{and}\quad q^{i,j = N_y} = -q^{i,j = N_y - 1} \, , \label{eq:no-flux-2}
\end{equation}
for all $i$.
The no-flux boundary conditions (\ref{eq:no-flux-2}) ensure the conservation of $\phi$ and $c$ exactly on the lattice.

\subsection{Spectral Method\label{sec:spectral}}

For the fluid velocity $\mathbf{u}$, we employ a spectral method based on a Fourier transform in the periodic $x$-direction 
and either a sine or a cosine transform in the $y$-direction.
We first compute the velocity field generated by the hydrodynamic force $\mathbf{f}$, before superimposing the imposed shear flow.

To satisfy the impermeability conditions 
\begin{equation}
u_y(x,0) = u_y(x,L_y) = 0 \, ,
\end{equation} 
we use a sine transform for $u_y$ in the $y$-direction.
For $u_x$, we impose free-slip boundary conditions at the walls
\begin{equation}
\partial_y u_x(x,0) = \partial_y u_x(x,L_y) = 0 \, ,
\end{equation}
and hence we use a cosine transform in the $y$-direction.
More precisely, we define the mixed-transforms as follows:
\begin{subequations} \label{eq:FT} 
\begin{align}
u_x(\mathbf{r}) &= \sum_{\mathbf{k}} \widetilde{u}_x(\mathbf{k}) e^{ik_xx}\cos(k_yy) \, , \\
u_y(\mathbf{r}) &= \sum_{\mathbf{k}} \widetilde{u}_y(\mathbf{k}) e^{ik_xx}\sin(k_yy) \, . 
\end{align} 
\end{subequations}
The summations are taken over the discrete set of wavevectors $\mathbf{k}=(k_x,k_y)$, where
\begin{equation}
k_x = \frac{2\pi n}{L_x} \, , \quad k_y = \frac{\pi m }{L_y} \, , 
\end{equation}
with $n\in\mathbb{Z}$. For the cosine expansion, $m=0,1,2,\dots$, whereas for the sine expansion, $m=1,2,3,\dots$.

With this choice of basis functions, the incompressibility condition (\ref{eq:incompressibility}) becomes
\begin{equation}
ik_x \widetilde{u}_x(\mathbf{k}) + k_y \widetilde{u}_y(\mathbf{k}) = 0 \, .
\end{equation}
Numerically, we first apply a fast Fourier transform (FFT) in the $x$-direction, followed by a discrete cosine transform (DCT) for $u_x$ and a discrete sine transform (DST) for $u_y$ in the $y$-direction, to obtain the transformed velocity field $\widetilde{\mathbf{u}}(\mathbf{k})$ from $\mathbf{u}(\mathbf{r})$.
Taking divergence of the Stokes equation (\ref{eq:Stokes}) and using the incompressibility condition (\ref{eq:incompressibility}), we obtain
\begin{equation}
\nabla^2P = \boldsymbol{\nabla}\cdot\mathbf{f} \, .
\end{equation}
For consistency with the mixed-transform representation of the velocity field, we expand $P(\mathbf{r})$ and $f_x(\mathbf{r})$ using cosine transforms, and $f_y(\mathbf{r})$ using a sine transform in the $y$-direction.

After transformation, the Stokes equation (\ref{eq:Stokes}) then becomes:
\begin{equation}
\begin{pmatrix}
\widetilde{u}_x \\
\widetilde{u}_y
\end{pmatrix}
=
\frac{1}{\eta k^4}
\begin{pmatrix}
k_y^2 & ik_xk_y \\
-ik_xk_y & k_x^2
\end{pmatrix}
\begin{pmatrix}
\widetilde{f}_x \\
\widetilde{f}_y
\end{pmatrix} \, ,
\end{equation}
for $\mathbf{k}\neq\boldsymbol{0}$, where $k^2=k_x^2+k_y^2$.
The uniform velocity mode $\widetilde{u}_x(\mathbf{0})$ is set to zero, assuming there is no net flow in the $x$-direction. 
The inverse transforms (\ref{eq:FT}) are then used to recover $\mathbf{u}(\mathbf{r})$ from $\widetilde{\mathbf{u}}(\mathbf{k})$.

To impose a shear rate $\dot{\gamma}=$ constant, we superimpose the simple shear solution onto the velocity field obtained above:
\begin{equation}
\mathbf{u}(\mathbf{r}) \rightarrow \mathbf{u}(\mathbf{r}) + \dot{\gamma}\left(y - \frac{L_y}{2} \right)\hat{\mathbf{x}} \, .
\end{equation}
The resulting velocity field satisfies incompressibility condition and the boundary conditions
\begin{subequations}
\begin{align}
u_y(x,0) &= u_y(x,L_y) = 0 \, ,\\
\partial_y u_x(x,0) &= \partial_y u _x(x,L_y) = \dot{\gamma} \, .
\end{align}
\end{subequations}
Physically, this corresponds to imposing a constant viscous shear stress at the walls, $\sigma_{xy}^\text{visc}=\eta\dot{\gamma}$.

We note that the present approach differs from the more established lattice Boltzmann method~\cite{Carenza2019}.
In lattice Boltzmann simulations, shear flow is typically imposed through moving-wall boundary conditions implemented via bounce-back schemes acting on the distribution functions, which are only first-order accurate in spatial discretisation~\cite{kruger2017lattice}.
Finally, the FFT algorithm scales as $(N_xN_y)\log(N_xN_y)$, whereas finite difference scheme scales as $N_xN_y$.
Thus for optimal numerical simulations, we use hybrid spectral on $\mathbf{u}$ and finite difference for the rest of the field variables.

\begin{figure}
    \centering
    \includegraphics[width=0.9\linewidth]{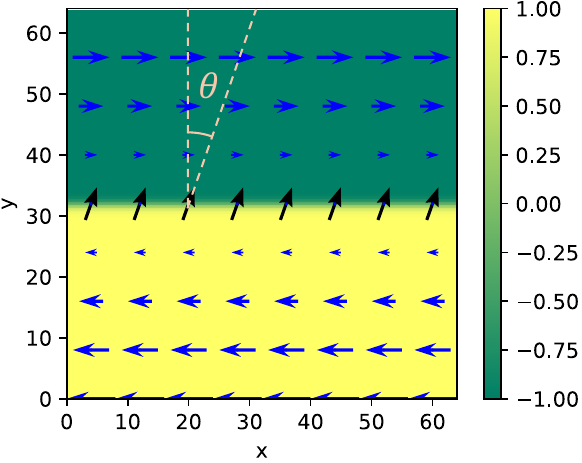}
    \caption{\justifying
    Snapshot of a simulation showing a flat interface located at $y=L_y/2=32$ under an imposed tangential shear flow in the $x$-direction.
    Black arrows denote the surfactant polarization (average orientation) $\mathbf{p}(\mathbf{r})$, while blue arrows represent the fluid velocity field $\mathbf{u}(\mathbf{r})$ in the steady state.
    The color map shows the value of the binary fluid order parameter $\phi(\mathbf{r})$, ranging from yellow ($\phi>0$) to green ($\phi<0$).
    The tilting angle $\theta$ is defined to be the angle between the polarization and the interface normal.
    In equilibrium ($\dot{\gamma}=0$), $\mathbf{p}$ will be perpendicular to the interface ($\theta=0$).
    Parameters used: $L_x=L_y=64$, $\dot{\gamma}=0.5$, $c_0=0.244$, $\varepsilon=1$, $M=0.1$ and $\Gamma_t=\Gamma_r=\beta=B=\eta=1$.}
    \label{fig:flat-interface}
\end{figure}

\begin{figure}
  \centering
  \includegraphics[width=0.8\linewidth]{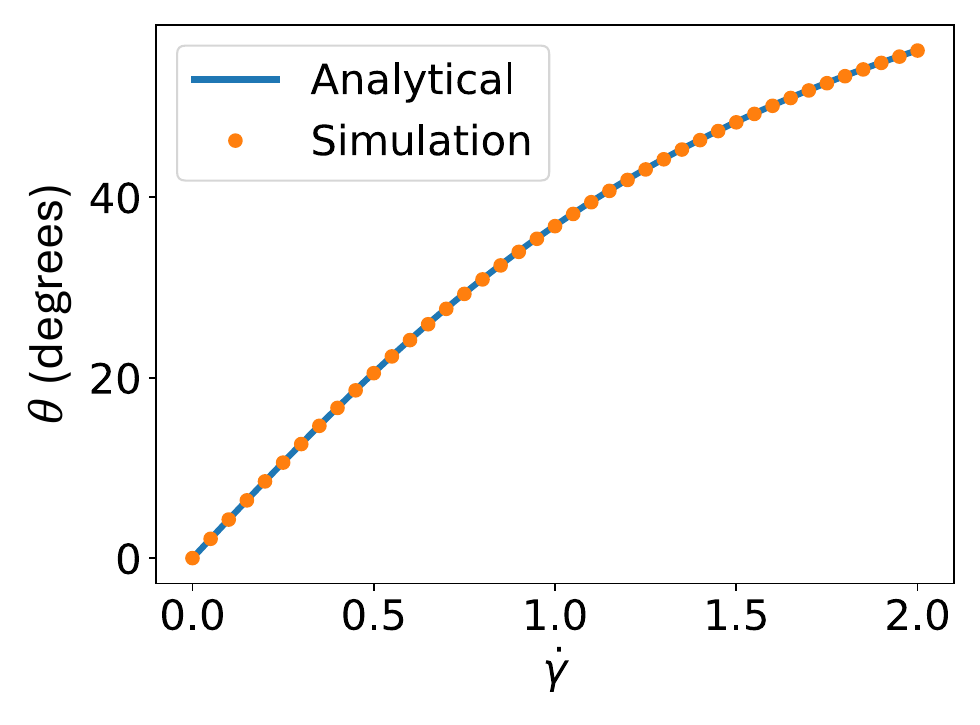}
  \caption{\justifying
  Tilting angle $\theta$ of the polarization field $\mathbf{p}$ relative to the interface normal under tangential shear flow as a function of the shear rate $\dot{\gamma}$. The orange points show the simulation results, while the blue curve represents the analytic solution obtained from perturbation theory. Parameters used: $c_0=0.244$, $\varepsilon=1$, $M=0.1$ and $\Gamma_t=\Gamma_r=\beta=B=\eta=1$.}
  \label{fig:tilting-angle}
\end{figure}

\section{Dynamic Surface Tension\label{sec:tilt}}

In this section, we examine how the effective surface tension of a surfactant-covered interface is modified in the presence of an imposed tangential shear flow. 
It is well known that, under equilibrium conditions, increasing the surfactant concentration reduces the effective surface tension. 
However, because surfactant molecules are anisotropic rather than spherical, they can tilt in response to an imposed shear flow; see Fig.~\ref{fig:flat-interface}. 
This tilting modifies their orientation relative to the interface and can counteract the surface-tension-reducing effect observed in the equilibrium case, where the molecules tend to align perpendicular to the interface.

To analyse the dependence of the effective surface tension $\sigma$ on both the average surfactant concentration $c_0$ and the shear rate $\dot{\gamma}$, we consider a flat interface located at $y=L_y/2$; see Fig.~\ref{fig:flat-interface}, with the oil phase ($\phi>0$) occupying the lower region $y<L_y/2$ and the water phase ($\phi<0$) occupying the upper region $y>L_y/2$.
In the absence of fluid flow ($\mathbf{u}=\boldsymbol{0}$), the surfactant concentration $c(\mathbf{r})$ and the polarization $\mathbf{p}(\mathbf{r})$ will relax toward an equilibrium state in which $c(y)$ exhibits a peak at $y=L_y/2$, indicating adsorption at the interface. At the same time, $\mathbf{p}(y)\propto \hat{\mathbf{y}}$ near the interface, showing that the surfactants tend to align perpendicular to the surface.

In the presence of an imposed tangential shear flow $\mathbf{u}=\dot{\gamma}(y-L_y/2)\hat{\mathbf{x}}$, the polarization $\mathbf{p}$ is no longer perpendicular to the interface but instead becomes tilted by an angle $\theta$ relative to the interface normal (see Fig.~\ref{fig:flat-interface}).
To quantify this tilting angle, we substitute the simple shear flow $\mathbf{u}=\dot{\gamma}(y-L_y/2)\hat{\mathbf{x}}$ into Eqs.~(\ref{eq:phidot}-\ref{eq:pdot}). 
By symmetry and in steady state, the out-of-plane component of the polarization vanishes, $p_z=0$.
The remaining fields $\phi$, $c$, $p_x$ and $p_y$ only depend on the coordinate $y$.  
In steady state, Eqs.~(\ref{eq:phidot}-\ref{eq:pdot}) then becomes
\begin{subequations}
\begin{align}
0 & =\frac{\partial^{2}}{\partial y^{2}}\left[-\beta\phi+\beta\phi^{3}-\frac{\beta}{2}\frac{\partial^{2}\phi}{\partial y^{2}}-\varepsilon\, \frac{\partial(cp_{y})}{\partial y}\right] , \\
0 & =\frac{\partial^{2}c}{\partial y^{2}}+\frac{\partial}{\partial y}\left[c\frac{\partial}{\partial y}\left(\frac{d}{2}\left(p_{x}^{2}+p_{y}^{2}\right)+\varepsilon p_{y}\frac{\partial\phi}{\partial y}\right)\right] , \\
0 & =\frac{1}{2}\left(\begin{array}{cc}
0 & \dot{\gamma}\\
-\dot{\gamma} & 0
\end{array}\right)\left(\begin{array}{c}
p_{x}\\
p_{y}
\end{array}\right)+\frac{Bd}{2(d+2)}\left(\begin{array}{cc}
0 & \dot{\gamma}\\
\dot{\gamma} & 0
\end{array}\right)\left(\begin{array}{c}
p_{x}\\
p_{y}
\end{array}\right) \nonumber\\
&\quad-\frac{d-1}{d\Gamma_{r}}\left[d\left(\begin{array}{c}
p_{x}\\
p_{y}
\end{array}\right)+\varepsilon\left(\begin{array}{c}
0\\
\partial\phi/\partial y
\end{array}\right)\right].
\end{align}
\end{subequations}
The resulting steady-state equations can then be solved perturbatively for small $\varepsilon$, where $\varepsilon$ characterizes the interaction strength between the surfactant molecules and the interface relative to the thermal energy scale $k_\mathrm{B} T$; see Section~\ref{sec:numerics}.
The perturbative solutions take the form
\begin{subequations} \label{eq:tilt-solution}
\begin{align}
\phi(y) &= -\tanh\left(y -\frac{L_y}{2} \right) + \mathcal{O}(\varepsilon^2) \, , \\
c(y) &= c_0 + \mathcal{O}(\varepsilon^2) \, , \\
p_x(y) &= \frac{\varepsilon A_+\lambda}{d(\lambda^2-A_+A_-)}\,\text{sech}^2\left(y -\frac{L_y}{2} \right)  + \mathcal{O}(\varepsilon^2) \, , \\
p_y(y) &= \frac{\varepsilon\lambda^2}{d(\lambda^2-A_+A_-)}\,\text{sech}^2\left(y -\frac{L_y}{2} \right)  + \mathcal{O}(\varepsilon^2) \, ,
\end{align}
\end{subequations}
where
\begin{equation}
\lambda = \frac{d-1}{\Gamma_r} \quad\text{and}\quad A_{\pm} = \frac{\dot{\gamma}}{2} \left( \frac{Bd}{d+2} \pm 1 \right) .
\end{equation}
Note that in our dimensionless units, $\Gamma_r=D_r^{-1}$ is the rotational diffusion timescale of the surfactants.
The tilting angle $\theta$ is then defined by $\tan\theta=p_x/p_y$.
We show in Fig.~\ref{fig:tilting-angle} the simulation results (orange points), which agree well with the perturbative solution (blue line).

\begin{figure}
    \centering
    \includegraphics[width=0.9\linewidth]{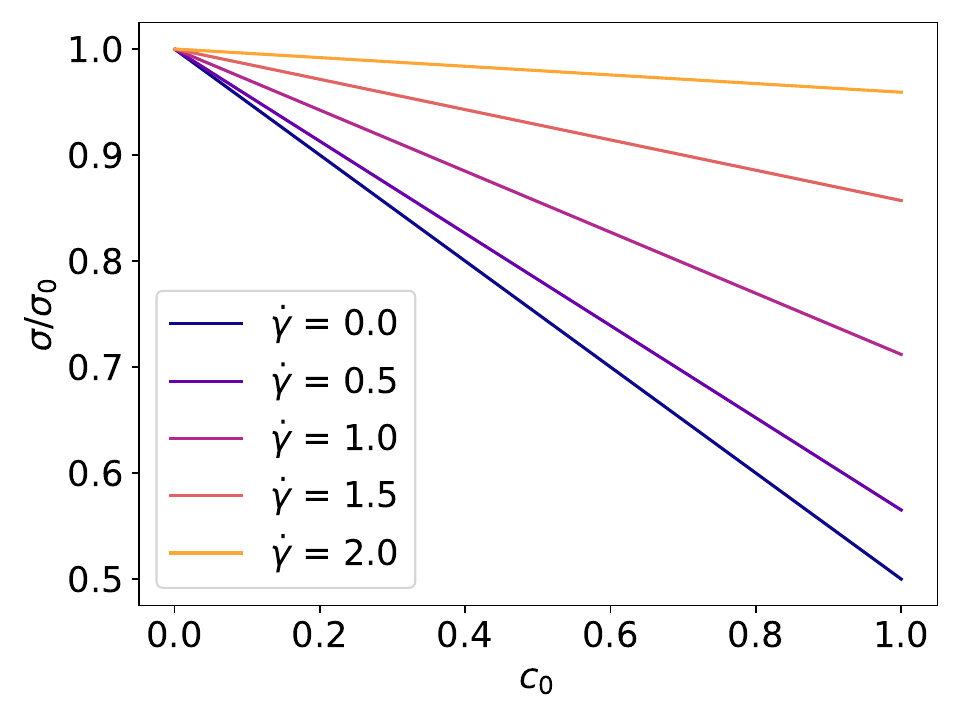}
    \caption{\justifying
    Effective surface tension $\sigma$ as a function of the average surfactant concentration $c_0$ for different shear rates $\dot{\gamma}$ [cf. Eq.~(\ref{eq:sheared-surface-tension})].
    As in the equilibrium case, the addition of surfactant lowers the effective surface tension, whereas an imposed tangential shear counteracts this reduction. Parameters used: $\varepsilon=\Gamma_r=1$.}
    \label{fig:sheared-surface-tension}
\end{figure}

Finally, using the steady–state solution (\ref{eq:tilt-solution}), the effective surface tension $\sigma(c_0,\dot{\gamma})$ can then be obtained from the excess free energy.
First, we define the grand potential to be:
\begin{equation}
\Phi[\phi,c,\mathbf{p}] = \mathcal{F}[\phi,c,\mathbf{p}] - \int \mu_c \, c(\mathbf{r}) \, d^dr =  \int w(\mathbf{r}) \, d^dr \, ,
\end{equation}
where $\mathcal{F}$ is the free energy defined in Eq.~(\ref{eq:free-energy-dimensionless}).
Mathematically, $\Phi$ is the Legendre transform of $\mathcal{F}$.
The quantity $\mu_c$ is the chemical potential of the surfactant reservoir, which is fixed as $\mu_c=\ln(c_0)+1$.
This choice ensures that the equilibrium condition $\delta\Phi/\delta c=0$ is satisfied.
Additionally, we define $w(\mathbf{r})$ to be the grand potential density.
The effective surface tension is then defined to be the excess grand potential across the interface:
\begin{equation} \label{eq:effective-surface-tension}
\sigma = \int_{-\infty}^{\infty} [w(\bar{y}) - w(\bar{y}\rightarrow\infty)] \, d\bar{y} \, ,
\end{equation}
where $\bar{y}=y-L_y/2$.
Note that the interface, assumed to be planar, is located at $y=L_y/2$.
Eq.~(\ref{eq:effective-surface-tension}) has also been shown to be equivalent to Gibbs adsorption isotherm~\cite{hardy2026kinetic}.

Substituting the perturbative steady state solution (\ref{eq:tilt-solution}) into (\ref{eq:effective-surface-tension}), 
we then obtain the effective surface tension as a function of average surfactant concentration $c_0$ and shear rate $\dot{\gamma}$:
\begin{equation}
    \sigma(c_0,\dot{\gamma}) = \sigma_0 - \frac{2c_0\lambda^2\left[ \lambda^2 - A_+(A_++2A_-) \right]}{3d(\lambda^2-A_+A_-)^2} \, \varepsilon^2 + \mathcal{O}(\varepsilon^4) \, , \label{eq:sheared-surface-tension}
\end{equation}
where $\sigma_0=2\beta/3$ is the bare surface tension in the absence of surfactant ($c_0=0$).
Note that $\lambda\propto\Gamma_r^{-1}\propto D_r$ and $A_\pm\propto\dot{\gamma}$. 
Thus, in the case of weak shear flow compared with the rotational diffusion constant, $\dot{\gamma}\ll D_r$, 
the effective surface tension~(\ref{eq:sheared-surface-tension}) reduces to
\begin{equation}
\sigma(c_0,\dot{\gamma}\ll D_r) = \sigma_0 - \frac{2c_0}{3d}\, \varepsilon^2 + \mathcal{O}(\varepsilon^4) \, . \label{eq:effective-sigma}
\end{equation}
In both cases, the surface tension decreases linearly with increasing surfactant concentration $c_0$.
Note that equations \eqref{eq:sheared-surface-tension} and \eqref{eq:effective-sigma} are derived for a general $d$-dimensional system, where we have an interface on the $x$-$z$ plane and shear flow on the $x$-$y$ plane.
However, in numerical simulations below, we restrict ourselves to $d=2$ dimensional case.

Fig.~\ref{fig:sheared-surface-tension} shows the effective surface tension $\sigma$ [Eq.~(\ref{eq:sheared-surface-tension})] as a function of the average surfactant concentration $c_0$ for different shear rates $\dot{\gamma}$. 
In the absence of shear, the addition of surfactant reduces the surface tension due to adsorption at the interface. 
However, as the shear rate increases, this reduction becomes weaker. 
Thus, shear can dynamically modify the effective surface tension.

To characterize the effect of shear-induced surfactant reorientation on the surface tension, and subsequently on the droplet dynamics considered in the next section, we introduce the dimensionless rotational P\'eclet number,
\begin{equation}
\text{Pe}_r = \frac{\dot{\gamma}}{D_r} \, . 
\end{equation}
For $\text{Pe}_r\ll 1$, rotational diffusion dominates over shear-induced reorientation, 
and the surfactant molecules remain approximately aligned with the interface normal. 
In this case, we may use the formula for the effective surface tension in Eq.~(\ref{eq:effective-sigma}).
By contrast, for $\text{Pe}_r\gg 1$, shear dominates, causing the surfactant orientations to deviate significantly from the interface normal. 
In this case, we do not have a simple expression for the macroscopic surface tension since Eq.~(\ref{eq:sheared-surface-tension}) depends on the local shear flow.

\begin{figure}
  \centering
  \includegraphics[width=0.7\linewidth]{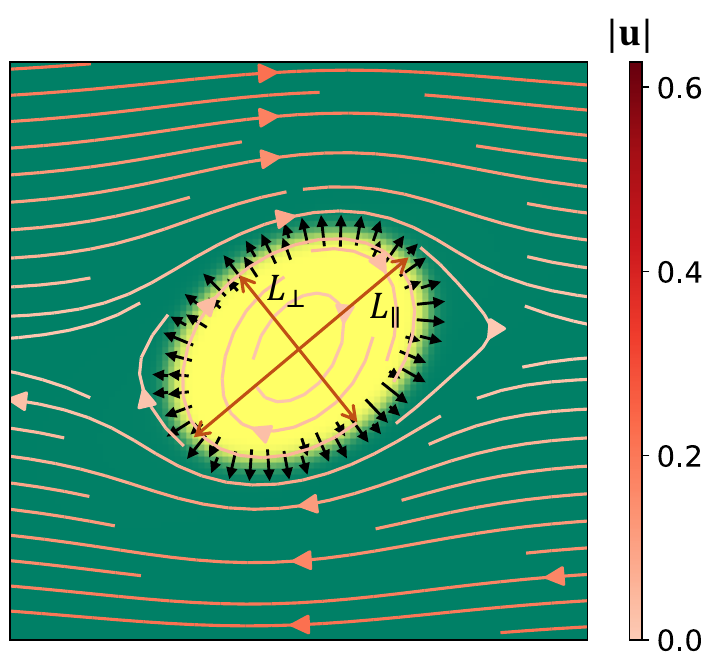}
  \caption{\justifying
  Snapshot of a deformed droplet under an imposed shear flow at steady state.
  The lines represent the streamlines of the fluid velocity field $\mathbf{u}(\mathbf{r})$, with the color scale indicating its magnitude $|\mathbf{u}|$.
  The black arrows indicate the polarization field $\mathbf{p}(\mathbf{r})$, or the average orientation of the surfactants. 
  The rotational P\'eclet number is $\text{Pe}_r=0.01$. Since $\text{Pe}_r\ll1$, $\mathbf{p}(\mathbf{r})$ are aligned perpendicular to the interface.
  The image shows the cropped region $(x,y)\in[0.31L_x,0.69L_x]\times[0.31L_y,0.69L_y]$.
  Other parameters used: $\dot{\gamma}=0.01$ and $c_0=0.244$.}
  \label{fig:sheared-droplet}
\end{figure}

\section{Droplet Deformation under Shear\label{sec: drop}}

We investigate the deformation of an initially circular droplet of radius $R$ under imposed shear, as illustrated in Fig.~\ref{fig:sheared-droplet}.
The domain is periodic in the $x$-direction. 
At the walls $y=0$ and $y=L_y$, we impose impermeability conditions, $u_y=0$, and prescribe the viscous shear stress
\begin{equation}
\sigma_{xy}^{\text{visc}}(x,0)=\sigma_{xy}^{\text{visc}}(x,L_y) = \eta\dot{\gamma} \, ,
\end{equation}
where $\dot{\gamma}$ is the imposed wall shear rate (see Sec.~\ref{sec:spectral} for details of the implementation).
Unless otherwise stated, we use $R=10$, $L_x=L_y=128$, $M=0.1$ and $\varepsilon=\beta=\eta=1$. 
The channel width is chosen to limit confinement effects, with $R/L_y=10/128$ for these parameters. 
We compare pure droplets with mean surfactant concentration $c_0=0$, and surfactant-covered droplets, with $c_0=0.244$.

The surfactant concentration is initialized to be uniform, $c(\mathbf{r},0)=c_0$, while the surfactant orientations are initially
isotropic, so that $\mathbf{p}(\mathbf{r},0)=\boldsymbol{0}$.
At steady state, $c(\mathbf{r},t\to\infty)$ peaks at the interface.
Likewise, $\mathbf{p}$ becomes polarized/nonzero near the interface and approaches zero in the bulk phases.

The droplet is initially centred at $y=L_y/2$. Under the imposed shear, it elongates, with its major axis inclined to the flow direction.
At sufficiently low shear rates, the droplet relaxes to a steady deformed shape. 
As the shear rate increases, the steady shape can lose stability, and the droplet may exhibit sustained oscillations, breakup, or more complex dynamics~\cite{Kaoui_PRE_2011,Kaoui_soft_2012}.

The extent of deformation is quantified by the deformation parameter:
\begin{equation}
  \text{D} = \frac{L_\parallel-L_\perp}{L_\parallel+L_\perp} \, , \label{eq:deformation}
\end{equation}
where $L_\parallel$ and $L_\perp$ are the lengths of the major and minor axes of the drop respectively (see Fig.~\ref{fig:sheared-droplet}). 
Numerically, the deformation parameter $D$ is computed from the inertia tensor:
\begin{equation}
  \underline{\underline{\mathbf{I}}} = \left( \begin{matrix}
   \sum_i{\Delta y_i^2}  & -\sum_i{\Delta x_i \Delta y_i}  \\ 
     -\sum_i{\Delta x_i \Delta y_i} & \sum_i{\Delta x_i^2}  \\
  \end{matrix} \right) ,
\end{equation}
where $\Delta x_i = x_i - x_\text{mean}$ and $\Delta y_i = y_i - y_\text{mean}$, and $x_i$ and $y_i$ are the spatial coordinates of pixels that are occupied by the droplet phase.
The droplet's center of mass is located at $(x_{\text{mean}},y_{\text{mean}})$. 
The square-root of the eigenvalues $\{\Lambda_{\text{max}},\Lambda_{\text{min}}\}$ of this matrix define the semi-axis lengths of the equivalent ellipse, namely $L_\parallel =N \sqrt{\Lambda_{\text{max}}}$ and $L_\perp =N \sqrt{\Lambda_\text{min}}$, where $\Lambda_{\text{max}}\ge\Lambda_{\text{min}}$ and $N$ is some constant.

Note that during simulation, the droplet may shrink slightly, so we define the average radius as
\begin{equation}
\bar{R} = \sqrt{\frac{\text{surface area of the ellipse}}{\pi}},
\end{equation}
which may not be necessarily equal to the original radius $R$.

We define the capillary number as the ratio of the imposed viscous stress
$\eta\dot{\gamma}$ to the reference capillary stress $\sigma_0/\bar{R}$,
\begin{equation}
\operatorname{Ca} = \frac{\eta\dot{\gamma}\bar{R}}{\sigma_0},
\label{eq:capillary}
\end{equation}
where $\eta$ is the common viscosity of the two fluid phases and
$\sigma_0$ is the bare interfacial tension in the limit $c_0\to0$.
We use $\sigma_0$ as the reference tension for both pure and
surfactant-covered droplets.

For each imposed shear rate $\dot{\gamma}$, we measure the steady-state deformation $\text{D}$ and the average radius $\bar{R}$.
We then plot the deformation parameter $\text{D}$ as a function of capillary number $\text{Ca}$.
We restrict our simulations to $0\leq\operatorname{Ca}<0.8$, over which the droplet remains intact.
We consider two regimes below.

\begin{figure}
    \includegraphics[width=1.0\linewidth]{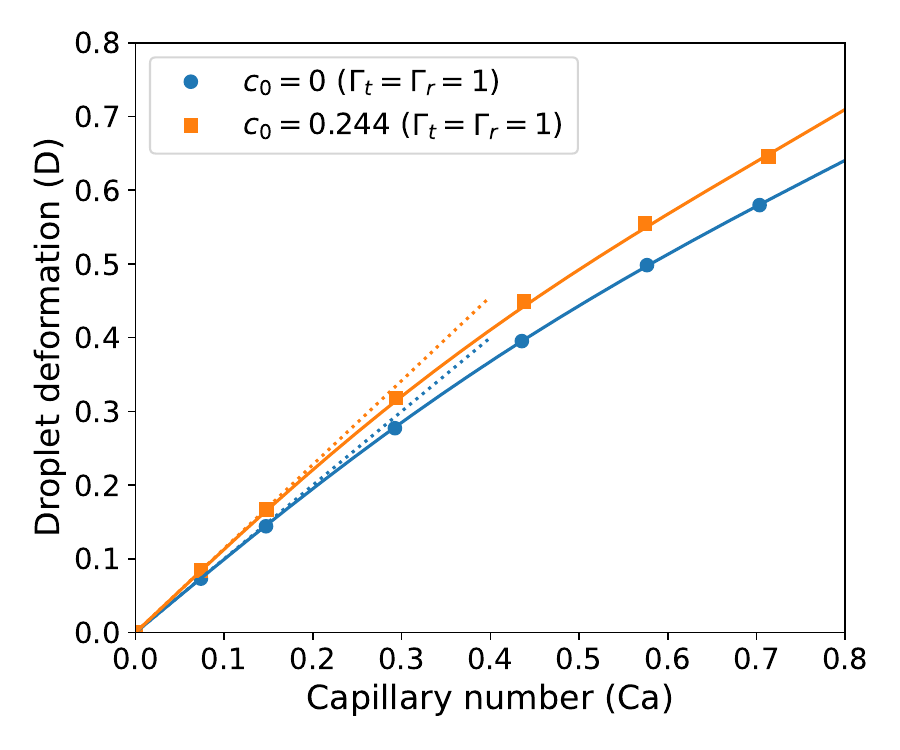}
    \caption{\justifying
    Steady-state deformation $\text{D}$ versus capillary number $\mathrm{Ca}$ for pure droplets ($c_0=0$, blue circles) and surfactant-covered droplets
($c_0=0.244$, orange squares). 
The dotted lines show Taylor's small-deformation predictions evaluated using the effective capillary number defined in Eq.~(\ref{eq:effective-Ca}).
The solid lines show the phenomenological Maffettone--Minale predictions of Eq.~(\ref{eq:MM-2}), evaluated using the same effective capillary number. 
Both curves are fitted using the same value, $\text{Ca}_{\text{x}}\simeq 0.552$. }
  \label{fig:unbounded}
\end{figure}

\subsection{$\text{Pe}_r$ remains small for all values of $\text{Ca}$}
 
As $\text{Ca}$ increases, $\text{Pe}_r$ also increases as both are proportional to the shear rate $\dot{\gamma}$.
In this subsection, we choose the rotational diffusion constant $D_r$ to be sufficiently large ($D_r=1$) so that $\text{Pe}_r$ remains small over the entire range $\text{Ca}\in[0,0.8]$ being considered.
In this regime, the surfactant molecules remain approximately aligned with the interface normal, and Eq.~(\ref{eq:effective-sigma}) may therefore be used for the effective surface tension.
To describe the steady-state droplet deformation $D$ as a function of the capillary number $\mathrm{Ca}$,
we extend the phenomenological Maffettone–Minale (MM) theory~\cite{megias2006determination} to describe the dependence of $D$ on $\mathrm{Ca}$ beyond the linear regime originally predicted by Taylor~\cite{taylor1934formation}.

We first consider a pure system without surfactants ($c_0=0$).
For sufficiently small capillary numbers, the deformation parameter is well described by Taylor’s result~\cite{taylor1934formation}, which has been reduced to two-dimension~\cite{9dgt-rkp5}:
\begin{align}
\text{D}^\text{T} = \text{Ca} \, , \quad \text{for small Ca}. \label{eq:Taylor} 
\end{align} 
(Note that we have equal viscosities inside and outside the droplet.)
At larger capillary numbers, this linear scaling breaks down, and a phenomenological MM theory has been proposed instead~\cite{megias2006determination}:
\begin{align}
  \text{D}^{\text{MM}} = \frac{f_2\text{Ca}}{\sqrt{f_1^2 + \text{Ca}^2} + \sqrt{f_1^2 + (1-f_2^2)\text{Ca}^2}} \, ,  \label{eq:MM}
\end{align}
where $f_1$ and $f_2$ are some phenomenological parameters.
In the limit $\text{Ca}\rightarrow0$, we should recover Taylor's result $\text{D}^{\text{MM}}=\text{Ca}$, which gives us a constraint between $f_1$ and $f_2$
\begin{equation}
f_2 = 2f_1 \, .
\end{equation}
We may then introduce a phenomenological fitting parameter $\text{Ca}_\text{x}$ such that $f_1=\text{Ca}_\text{x}$ and $f_2=2\text{Ca}_\text{x}$.
Eq.~(\ref{eq:MM}) can then be written as
\begin{equation}
\text{D}^{\text{MM}} = \frac{2\text{Ca}_\text{x}\text{Ca}}{\sqrt{\text{Ca}_\text{x}^2 + \text{Ca}^2} + \sqrt{\text{Ca}_\text{x}^2 + (1-4\text{Ca}_\text{x}^2)\text{Ca}^2}} \, . \label{eq:MM-2}
\end{equation}
Fitting the phenomenological model to the pure-droplet simulation data (blue circles in Fig.~\ref{fig:unbounded}) yields
$\mathrm{Ca}_{\mathrm{x}}\simeq 0.552$. 
The resulting curve (blue line) closely matches the measured deformation over the full range of capillary numbers shown, using $\mathrm{Ca}_{\mathrm{x}}$ as the sole fitting parameter.

In the presence of surfactant, it is sufficient to replace the capillary number in Eqs.~(\ref{eq:MM-2}) with the effective capillary number:
\begin{equation}
\text{Ca} \rightarrow \frac{\sigma_0}{\sigma} \, \text{Ca}  \, , \quad\text{with}\quad
\sigma = \sigma_0 - \frac{2c_0}{6}\varepsilon^2 \,. \label{eq:effective-Ca}
\end{equation}
where $\sigma$ is the reduced interfacial tension due to surfactant adsorption [cf. Eq.~(\ref{eq:effective-sigma})].
This prescription assumes that the average surfactant orientation $\mathbf{p}$ remains approximately perpendicular to the interface.

The orange squares in Fig.~\ref{fig:unbounded} show the steady-state deformation $D$ of surfactant-covered droplets ($c_0=0.244$) as a function
of the capillary number $\text{Ca}$. 
The solid orange line shows the phenomenological Maffettone--Minale prediction evaluated using the effective capillary number defined in Eq.~(\ref{eq:effective-Ca}). 
We use the same fitting value $\text{Ca}_{\text{x}}\simeq0.552$ obtained from the pure-droplet fit.
The resulting curve agrees closely with the simulation data across the range shown.

\begin{figure}
  \centering
  \includegraphics[width=0.8\linewidth]{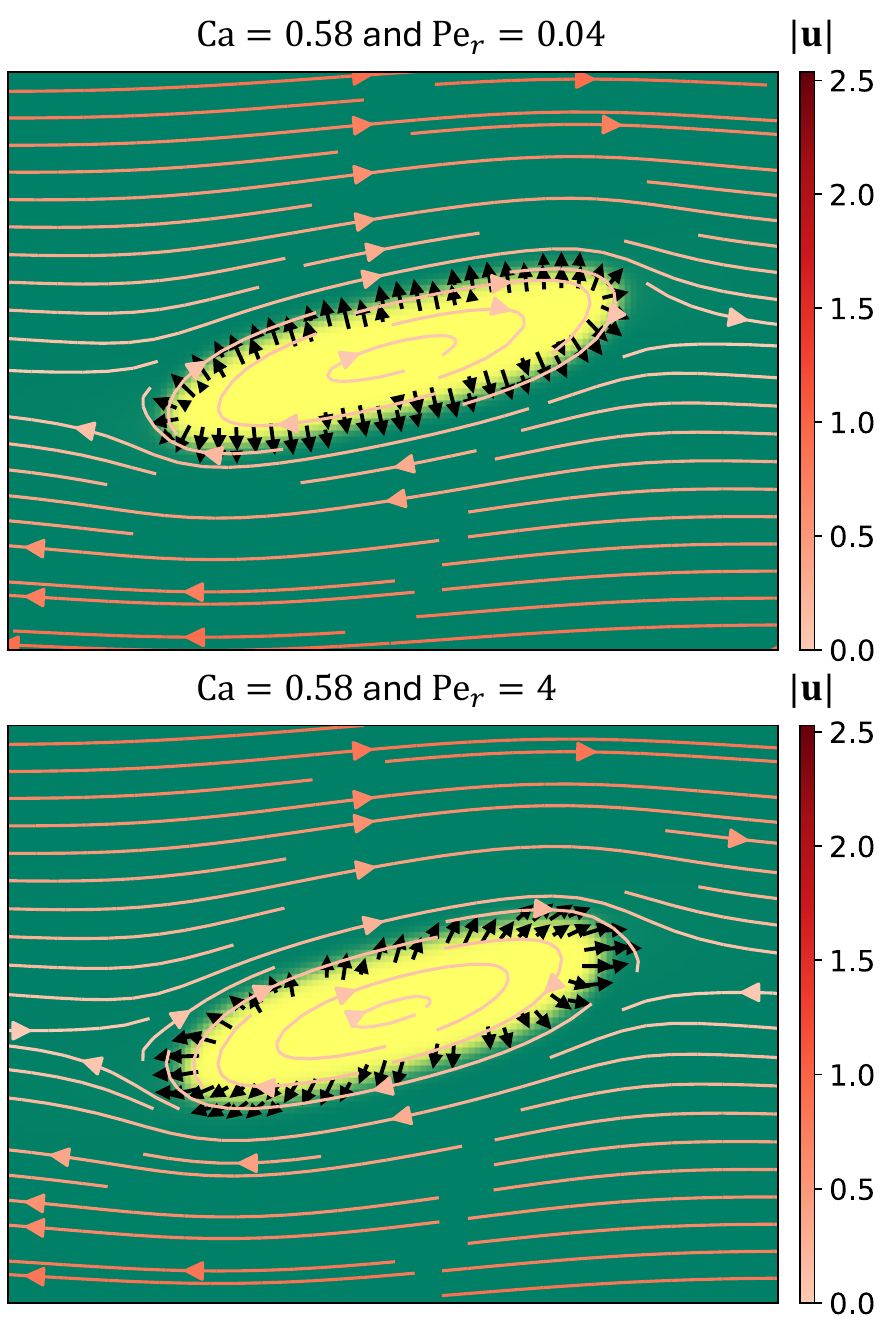}
  \caption{\justifying
  Two snapshots of a deformed droplet in the steady state at the same capillary number, $\text{Ca}=0.58$, but for two different values of the rotational Péclet number: $\text{Pe}_r=0.04$ (top) and $\text{Pe}_r=4$ (bottom).
  For $\text{Pe}_r=0.04$ (top), the average surfactant orientation $\mathbf{p}(\mathbf{r})$ (black arrows) remains approximately perpendicular to the interface. By contrast, for $\mathrm{Pe}_r=4$ (bottom), $\mathbf{p}(\mathbf{r})$ deviates significantly from the local interface normal.
  Both images show the cropped region $(x,y)\in[0.25L_x,0.75L_x]\times[0.31L_y,0.69L_y]$.
  Other parameters: $\dot{\gamma}=0.04$, $\Gamma_t=\Gamma_r=1$ (top) and $\Gamma_t=\Gamma_r=100$ (bottom).}
  \label{fig:sheared-droplet-large-Pe}
\end{figure}

\subsection{$\text{Pe}_r$ becomes large for some values of $\text{Ca}$}

In the previous subsection, we chose $D_r$ to be sufficiently large ($D_r=1$) so that $\text{Pe}_r$ remained small over the entire range of $\text{Ca}$ considered. 
Here, we instead choose a much smaller value ($D_r=\Gamma_r^{-1}=0.01$), such that $\text{Pe}_r$ is larger than unity for some values of $\text{Ca}$.
For instance, Fig.~\ref{fig:sheared-droplet-large-Pe} shows the steady state droplet deformation at $\text{Ca}=0.58$ but with two different values of $D_r$.
Fig.~\ref{fig:sheared-droplet-large-Pe} (top) with $D_r=1$ gives $\text{Pe}_r=0.04$, whereas Fig.~\ref{fig:sheared-droplet-large-Pe} (bottom) with $D_r=0.01$ gives $\text{Pe}_r=4$.
In the latter regime, the average surfactant orientation deviates significantly from the interface normal, and no simple analytical expression for the macroscopic effective surface tension is available.

\begin{figure}
  \centering
  \includegraphics[width=1.0\linewidth]{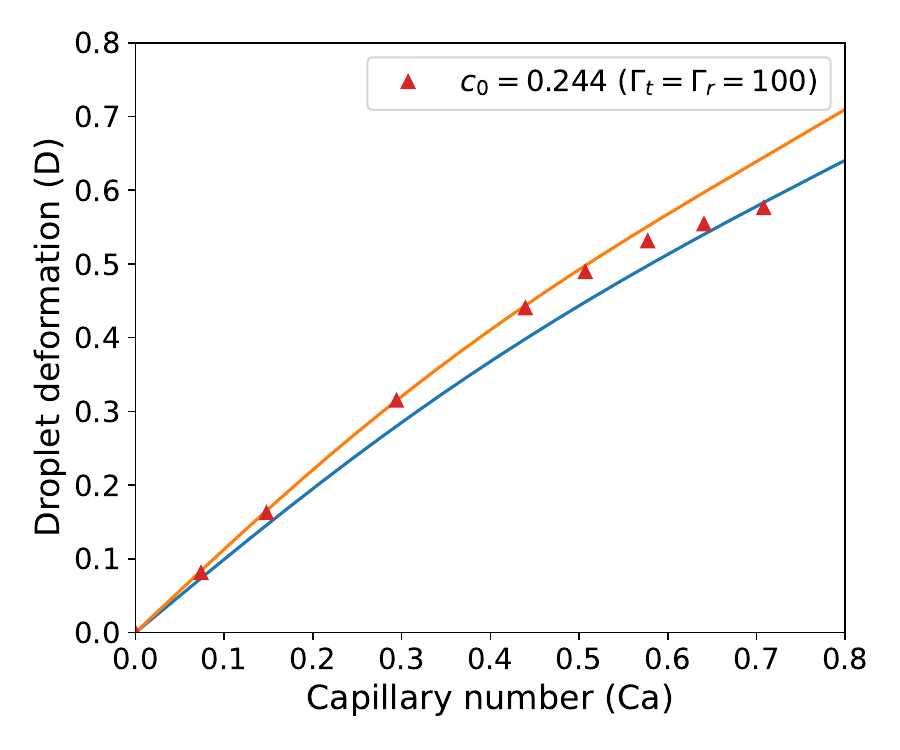}
  \caption{\justifying
   Steady-state deformation parameter $\text{D}$ as a function of the capillary number $\text{Ca}$ for a much smaller rotational diffusion constant $D_r=\Gamma_r^{-1}=0.01$. 
   The numerical results are shown as red triangles.
   The blue line shows the Maffettone--Minale (MM) prediction for pure droplets, and the orange line shows the prediction for surfactant-covered droplets.
At small $\text{Ca}$, the rotational P\'eclet number $\mathrm{Pe}_r$ is small, and $\text{D}$ follows the surfactant-covered prediction (orange line). 
As $\text{Ca}$ increases, shear reorients the surfactants, weakening the reduction in surface tension. 
The measured deformation then approaches the MM prediction for a pure droplet (blue line).
  \label{fig:large-Pe}}
\end{figure}

To study the effect of rotational P\'eclet number on droplet deformation, we plot the steady-state deformation parameter $\text{D}$ as a function of the capillary number $\text{Ca}$; see Fig.~\ref{fig:large-Pe}. 
Here, we use a much smaller rotational diffusion constant $D_r=0.01$ than in the previous subsection, so that the rotational P\'eclet number $\text{Pe}_r$ increases appreciably with $\text{Ca}$.

For small $\text{Ca}$, $\text{Pe}_r$ remains small, and rotational diffusion keeps the average surfactant orientation approximately normal to the interface. 
The deformation parameter therefore follows the MM prediction for a surfactant-covered droplet derived in the previous subsection (indicated orange line in Fig.~\ref{fig:large-Pe}). 
As $\text{Ca}$ increases, however, $\text{Pe}_r$ becomes of order unity or larger, and the imposed shear flow significantly reorients the surfactants away from the interface normal. 
This reorientation counteracts the surfactant-induced reduction in the effective surface tension.
Consequently, the droplet behaves increasingly like a pure droplet, and the measured deformation parameter moves towards the MM prediction for a pure droplet (blue line in Fig.~\ref{fig:large-Pe}). 

For a no-slip spherical particle of radius $a$, the ratio of the translational and rotational diffusion coefficients in physical units is $D_t/D_r=4a^2/3$. Using the interfacial width $\xi$ as the unit of length, this ratio becomes 
$D_t/D_r=\Gamma_r/\Gamma_t=(4/3)(a/\xi)^2$ in our dimensionless units. 
Physically, $a$ has to be at least smaller than $\xi$.
Here, we $\Gamma_t=\Gamma_r$, corresponding to $a/\xi=\sqrt{3}/2$.

The typical size of a surfactant molecule is of order $1\,\text{nm}$.
In water, this corresponds to a rotational diffusion constant of approximately $D_r\sim10^{8}\,\text{s}^{-1}$.
The shear-induced reorientation effects shown in Fig.~\ref{fig:large-Pe} are unlikely to be significant in typical microfluidic experiments and may become observable only at much larger shear rates accessible in nanofluidic systems. 
A more experimentally accessible alternative may be provided by colloidal particles adsorbed at fluid--fluid interfaces, as in Pickering emulsions. 
Such particles can play a role analogous to that of molecular surfactants while exhibiting much slower rotational diffusion, making shear-induced reorientation easier to observe.

\section{Conclusions}

In this work, we investigated how surfactant orientation influences effective surface tension and droplet deformation under shear flow. 
Using a phase-field model that incorporates both surfactant concentration and polarization, we showed that tangential shear tilts the surfactant orientation away from the interface normal, reducing its ability to lower the interfacial free energy. 

The competition between shear-induced reorientation and rotational diffusion is characterised by the rotational P\'eclet number, $\text{Pe}_r$. 
For small $\text{Pe}_r$, the surfactants remain approximately normal to the interface, allowing a simple expression for the surfactant-renormalized surface tension. 
This leads to modified Taylor and Maffettone--Minale predictions for small and large capillary numbers, respectively, both in good agreement with numerical simulations. 
At larger $\text{Pe}_r$, flow strongly distorts the surfactant orientation and progressively counteracts the reduction in surface tension, causing the steady-state deformation to approach that of a pure droplet.

Our results establish a direct link between microscopic surfactant reorientation and macroscopic interfacial behaviour. 
Natural extensions include chemically active surfactants or active stresses, for which the coupling between polarization, flow, and interfacial deformation may generate spontaneous droplet propulsion, rotation, breakup, and activity-driven morphological transitions~\cite{C4SM00550C,PhysRevResearch.2.032024,Bonelli2019,PhysRevX.11.021001}.

\begin{acknowledgments}
A.J.H.\ acknowledges EPSRC DTP studentship no. 2739112. E.T.\ acknowledges funding from EPSRC grant no.\ EP/W027194/1.
\end{acknowledgments}



\section*{Conflict of interest}

The author declares no conflict of interest.

\section*{Data Availability Statement}

The data supporting the results reported in the manuscript are available from the corresponding authors upon reasonable request.

\bibliography{main-revised}

\begin{thebibliography}{46}%
\makeatletter
\providecommand \@ifxundefined [1]{%
 \@ifx{#1\undefined}
}%
\providecommand \@ifnum [1]{%
 \ifnum #1\expandafter \@firstoftwo
 \else \expandafter \@secondoftwo
 \fi
}%
\providecommand \@ifx [1]{%
 \ifx #1\expandafter \@firstoftwo
 \else \expandafter \@secondoftwo
 \fi
}%
\providecommand \natexlab [1]{#1}%
\providecommand \enquote  [1]{``#1''}%
\providecommand \bibnamefont  [1]{#1}%
\providecommand \bibfnamefont [1]{#1}%
\providecommand \citenamefont [1]{#1}%
\providecommand \href@noop [0]{\@secondoftwo}%
\providecommand \href [0]{\begingroup \@sanitize@url \@href}%
\providecommand \@href[1]{\@@startlink{#1}\@@href}%
\providecommand \@@href[1]{\endgroup#1\@@endlink}%
\providecommand \@sanitize@url [0]{\catcode `\\12\catcode `\$12\catcode
  `\&12\catcode `\#12\catcode `\^12\catcode `\_12\catcode `\%12\relax}%
\providecommand \@@startlink[1]{}%
\providecommand \@@endlink[0]{}%
\providecommand \url  [0]{\begingroup\@sanitize@url \@url }%
\providecommand \@url [1]{\endgroup\@href {#1}{\urlprefix }}%
\providecommand \urlprefix  [0]{URL }%
\providecommand \Eprint [0]{\href }%
\providecommand \doibase [0]{https://doi.org/}%
\providecommand \selectlanguage [0]{\@gobble}%
\providecommand \bibinfo  [0]{\@secondoftwo}%
\providecommand \bibfield  [0]{\@secondoftwo}%
\providecommand \translation [1]{[#1]}%
\providecommand \BibitemOpen [0]{}%
\providecommand \bibitemStop [0]{}%
\providecommand \bibitemNoStop [0]{.\EOS\space}%
\providecommand \EOS [0]{\spacefactor3000\relax}%
\providecommand \BibitemShut  [1]{\csname bibitem#1\endcsname}%
\let\auto@bib@innerbib\@empty
\bibitem [{\citenamefont {Mulqueen}\ and\ \citenamefont
  {Blankschtein}(2002)}]{mulqueen2002theoretical}%
  \BibitemOpen
  \bibfield  {author} {\bibinfo {author} {\bibfnamefont {M.}~\bibnamefont
  {Mulqueen}}\ and\ \bibinfo {author} {\bibfnamefont {D.}~\bibnamefont
  {Blankschtein}},\ }\bibfield  {title} {\enquote {\bibinfo {title}
  {Theoretical and experimental investigation of the equilibrium oil- water
  interfacial tensions of solutions containing surfactant mixtures},}\
  }\href@noop {} {\bibfield  {journal} {\bibinfo  {journal} {Langmuir}\
  }\textbf {\bibinfo {volume} {18}},\ \bibinfo {pages} {365--376} (\bibinfo
  {year} {2002})}\BibitemShut {NoStop}%
\bibitem [{\citenamefont {M{\"o}bius}, \citenamefont {Miller},\ and\
  \citenamefont {Fainerman}(2001)}]{mobius2001surfactants}%
  \BibitemOpen
  \bibfield  {author} {\bibinfo {author} {\bibfnamefont {D.}~\bibnamefont
  {M{\"o}bius}}, \bibinfo {author} {\bibfnamefont {R.}~\bibnamefont {Miller}},\
  and\ \bibinfo {author} {\bibfnamefont {V.~B.}\ \bibnamefont {Fainerman}},\
  }\href@noop {} {\emph {\bibinfo {title} {Surfactants: chemistry, interfacial
  properties, applications}}},\ Vol.~\bibinfo {volume} {13}\ (\bibinfo
  {publisher} {Elsevier},\ \bibinfo {year} {2001})\BibitemShut {NoStop}%
\bibitem [{\citenamefont {Tadros}(2006)}]{tadros2006applied}%
  \BibitemOpen
  \bibfield  {author} {\bibinfo {author} {\bibfnamefont {T.~F.}\ \bibnamefont
  {Tadros}},\ }\href@noop {} {\emph {\bibinfo {title} {Applied surfactants:
  principles and applications}}}\ (\bibinfo  {publisher} {John Wiley \& Sons},\
  \bibinfo {year} {2006})\BibitemShut {NoStop}%
\bibitem [{\citenamefont {Schramm}\ and\ \citenamefont
  {Schramm}(2000)}]{schramm2000surfactants}%
  \BibitemOpen
  \bibfield  {author} {\bibinfo {author} {\bibfnamefont {L.~L.}\ \bibnamefont
  {Schramm}}\ and\ \bibinfo {author} {\bibfnamefont {L.~L.}\ \bibnamefont
  {Schramm}},\ }\href@noop {} {\emph {\bibinfo {title} {Surfactants:
  fundamentals and applications in the petroleum industry}}}\ (\bibinfo
  {publisher} {Cambridge university press},\ \bibinfo {year}
  {2000})\BibitemShut {NoStop}%
\bibitem [{\citenamefont {Shaban}, \citenamefont {Kang},\ and\ \citenamefont
  {Kim}(2020)}]{shaban2020surfactants}%
  \BibitemOpen
  \bibfield  {author} {\bibinfo {author} {\bibfnamefont {S.~M.}\ \bibnamefont
  {Shaban}}, \bibinfo {author} {\bibfnamefont {J.}~\bibnamefont {Kang}},\ and\
  \bibinfo {author} {\bibfnamefont {D.-H.}\ \bibnamefont {Kim}},\ }\bibfield
  {title} {\enquote {\bibinfo {title} {Surfactants: Recent advances and their
  applications},}\ }\href@noop {} {\bibfield  {journal} {\bibinfo  {journal}
  {Composites communications}\ }\textbf {\bibinfo {volume} {22}},\ \bibinfo
  {pages} {100537} (\bibinfo {year} {2020})}\BibitemShut {NoStop}%
\bibitem [{\citenamefont {Singh}\ and\ \citenamefont
  {Cameotra}(2004)}]{singh2004potential}%
  \BibitemOpen
  \bibfield  {author} {\bibinfo {author} {\bibfnamefont {P.}~\bibnamefont
  {Singh}}\ and\ \bibinfo {author} {\bibfnamefont {S.~S.}\ \bibnamefont
  {Cameotra}},\ }\bibfield  {title} {\enquote {\bibinfo {title} {Potential
  applications of microbial surfactants in biomedical sciences},}\ }\href@noop
  {} {\bibfield  {journal} {\bibinfo  {journal} {TRENDS in Biotechnology}\
  }\textbf {\bibinfo {volume} {22}},\ \bibinfo {pages} {142--146} (\bibinfo
  {year} {2004})}\BibitemShut {NoStop}%
\bibitem [{\citenamefont {Nitschke}\ and\ \citenamefont
  {Silva}(2018)}]{nitschke2018recent}%
  \BibitemOpen
  \bibfield  {author} {\bibinfo {author} {\bibfnamefont {M.}~\bibnamefont
  {Nitschke}}\ and\ \bibinfo {author} {\bibfnamefont {S.~S.~e.}\ \bibnamefont
  {Silva}},\ }\bibfield  {title} {\enquote {\bibinfo {title} {Recent food
  applications of microbial surfactants},}\ }\href@noop {} {\bibfield
  {journal} {\bibinfo  {journal} {Critical reviews in food science and
  nutrition}\ }\textbf {\bibinfo {volume} {58}},\ \bibinfo {pages} {631--638}
  (\bibinfo {year} {2018})}\BibitemShut {NoStop}%
\bibitem [{\citenamefont {Scamehorn}(1986)}]{scamehorn1986overview}%
  \BibitemOpen
  \bibfield  {author} {\bibinfo {author} {\bibfnamefont {J.~F.}\ \bibnamefont
  {Scamehorn}},\ }\href@noop {} {\emph {\bibinfo {title} {An overview of
  phenomena involving surfactant mixtures}}}\ (\bibinfo  {publisher} {ACS
  Publications},\ \bibinfo {year} {1986})\BibitemShut {NoStop}%
\bibitem [{\citenamefont {Rosen}\ and\ \citenamefont
  {Kunjappu}(2012)}]{rosen2012surfactants}%
  \BibitemOpen
  \bibfield  {author} {\bibinfo {author} {\bibfnamefont {M.~J.}\ \bibnamefont
  {Rosen}}\ and\ \bibinfo {author} {\bibfnamefont {J.~T.}\ \bibnamefont
  {Kunjappu}},\ }\href@noop {} {\emph {\bibinfo {title} {Surfactants and
  interfacial phenomena}}}\ (\bibinfo  {publisher} {John Wiley \& Sons},\
  \bibinfo {year} {2012})\BibitemShut {NoStop}%
\bibitem [{\citenamefont {Dave}\ and\ \citenamefont
  {Joshi}(2017)}]{dave2017concise}%
  \BibitemOpen
  \bibfield  {author} {\bibinfo {author} {\bibfnamefont {N.}~\bibnamefont
  {Dave}}\ and\ \bibinfo {author} {\bibfnamefont {T.}~\bibnamefont {Joshi}},\
  }\bibfield  {title} {\enquote {\bibinfo {title} {A concise review on
  surfactants and its significance},}\ }\href@noop {} {\bibfield  {journal}
  {\bibinfo  {journal} {Int. J. Appl. Chem}\ }\textbf {\bibinfo {volume}
  {13}},\ \bibinfo {pages} {663--672} (\bibinfo {year} {2017})}\BibitemShut
  {NoStop}%
\bibitem [{\citenamefont {Hardy}, \citenamefont {Daddi-Moussa-Ider},\ and\
  \citenamefont {Tjhung}(2024)}]{hardy2024hybrid}%
  \BibitemOpen
  \bibfield  {author} {\bibinfo {author} {\bibfnamefont {A.~J.}\ \bibnamefont
  {Hardy}}, \bibinfo {author} {\bibfnamefont {A.}~\bibnamefont
  {Daddi-Moussa-Ider}},\ and\ \bibinfo {author} {\bibfnamefont
  {E.}~\bibnamefont {Tjhung}},\ }\bibfield  {title} {\enquote {\bibinfo {title}
  {Hybrid particle-phase field model and renormalized surface tension in dilute
  suspensions of nanoparticles},}\ }\href@noop {} {\bibfield  {journal}
  {\bibinfo  {journal} {Physical Review E}\ }\textbf {\bibinfo {volume}
  {110}},\ \bibinfo {pages} {044606} (\bibinfo {year} {2024})}\BibitemShut
  {NoStop}%
\bibitem [{\citenamefont {Hardy}\ \emph {et~al.}(2026)\citenamefont {Hardy},
  \citenamefont {Cameron}, \citenamefont {McDonald}, \citenamefont
  {Daddi-Moussa-Ider},\ and\ \citenamefont {Tjhung}}]{hardy2026kinetic}%
  \BibitemOpen
  \bibfield  {author} {\bibinfo {author} {\bibfnamefont {A.~J.}\ \bibnamefont
  {Hardy}}, \bibinfo {author} {\bibfnamefont {S.}~\bibnamefont {Cameron}},
  \bibinfo {author} {\bibfnamefont {S.}~\bibnamefont {McDonald}}, \bibinfo
  {author} {\bibfnamefont {A.}~\bibnamefont {Daddi-Moussa-Ider}},\ and\
  \bibinfo {author} {\bibfnamefont {E.}~\bibnamefont {Tjhung}},\ }\bibfield
  {title} {\enquote {\bibinfo {title} {Kinetic theory of binary
  fluid–surfactant systems: A variational framework},}\ }\href
  {https://doi.org/10.1017/jfm.2026.11248} {\bibfield  {journal} {\bibinfo
  {journal} {Journal of Fluid Mechanics}\ }\textbf {\bibinfo {volume} {1029}},\
  \bibinfo {pages} {A54} (\bibinfo {year} {2026})}\BibitemShut {NoStop}%
\bibitem [{\citenamefont {Taylor}(1934)}]{taylor1934formation}%
  \BibitemOpen
  \bibfield  {author} {\bibinfo {author} {\bibfnamefont {G.~I.}\ \bibnamefont
  {Taylor}},\ }\bibfield  {title} {\enquote {\bibinfo {title} {The formation of
  emulsions in definable fields of flow},}\ }\href@noop {} {\bibfield
  {journal} {\bibinfo  {journal} {Proceedings of the Royal Society of London.
  Series A, containing papers of a mathematical and physical character}\
  }\textbf {\bibinfo {volume} {146}},\ \bibinfo {pages} {501--523} (\bibinfo
  {year} {1934})}\BibitemShut {NoStop}%
\bibitem [{\citenamefont {Maffettone}\ and\ \citenamefont
  {Minale}(1998)}]{maffettone1998equation}%
  \BibitemOpen
  \bibfield  {author} {\bibinfo {author} {\bibfnamefont {P.}~\bibnamefont
  {Maffettone}}\ and\ \bibinfo {author} {\bibfnamefont {M.}~\bibnamefont
  {Minale}},\ }\bibfield  {title} {\enquote {\bibinfo {title} {Equation of
  change for ellipsoidal drops in viscous flow},}\ }\href@noop {} {\bibfield
  {journal} {\bibinfo  {journal} {Journal of non-Newtonian fluid mechanics}\
  }\textbf {\bibinfo {volume} {78}},\ \bibinfo {pages} {227--241} (\bibinfo
  {year} {1998})}\BibitemShut {NoStop}%
\bibitem [{\citenamefont {Minale}(2010)}]{minale2010models}%
  \BibitemOpen
  \bibfield  {author} {\bibinfo {author} {\bibfnamefont {M.}~\bibnamefont
  {Minale}},\ }\bibfield  {title} {\enquote {\bibinfo {title} {Models for the
  deformation of a single ellipsoidal drop: a review},}\ }\href@noop {}
  {\bibfield  {journal} {\bibinfo  {journal} {Rheologica acta}\ }\textbf
  {\bibinfo {volume} {49}},\ \bibinfo {pages} {789--806} (\bibinfo {year}
  {2010})}\BibitemShut {NoStop}%
\bibitem [{\citenamefont {Sibillo}\ \emph {et~al.}(2006)\citenamefont
  {Sibillo}, \citenamefont {Pasquariello}, \citenamefont {Simeone},
  \citenamefont {Cristini},\ and\ \citenamefont {Guido}}]{sibillo2006drop}%
  \BibitemOpen
  \bibfield  {author} {\bibinfo {author} {\bibfnamefont {V.}~\bibnamefont
  {Sibillo}}, \bibinfo {author} {\bibfnamefont {G.}~\bibnamefont
  {Pasquariello}}, \bibinfo {author} {\bibfnamefont {M.}~\bibnamefont
  {Simeone}}, \bibinfo {author} {\bibfnamefont {V.}~\bibnamefont {Cristini}},\
  and\ \bibinfo {author} {\bibfnamefont {S.}~\bibnamefont {Guido}},\ }\bibfield
   {title} {\enquote {\bibinfo {title} {Drop deformation in microconfined shear
  flow},}\ }\href@noop {} {\bibfield  {journal} {\bibinfo  {journal} {Physical
  review letters}\ }\textbf {\bibinfo {volume} {97}},\ \bibinfo {pages}
  {054502} (\bibinfo {year} {2006})}\BibitemShut {NoStop}%
\bibitem [{\citenamefont {Feigl}\ \emph {et~al.}(2003)\citenamefont {Feigl},
  \citenamefont {Kaufmann}, \citenamefont {Fischer},\ and\ \citenamefont
  {Windhab}}]{feigl2003numerical}%
  \BibitemOpen
  \bibfield  {author} {\bibinfo {author} {\bibfnamefont {K.}~\bibnamefont
  {Feigl}}, \bibinfo {author} {\bibfnamefont {S.~F.}\ \bibnamefont {Kaufmann}},
  \bibinfo {author} {\bibfnamefont {P.}~\bibnamefont {Fischer}},\ and\ \bibinfo
  {author} {\bibfnamefont {E.~J.}\ \bibnamefont {Windhab}},\ }\bibfield
  {title} {\enquote {\bibinfo {title} {A numerical procedure for calculating
  droplet deformation in dispersing flows and experimental verification},}\
  }\href@noop {} {\bibfield  {journal} {\bibinfo  {journal} {Chemical
  Engineering Science}\ }\textbf {\bibinfo {volume} {58}},\ \bibinfo {pages}
  {2351--2363} (\bibinfo {year} {2003})}\BibitemShut {NoStop}%
\bibitem [{\citenamefont {Cristini}\ \emph {et~al.}(2003)\citenamefont
  {Cristini}, \citenamefont {Guido}, \citenamefont {Alfani}, \citenamefont
  {B{\l}awzdziewicz},\ and\ \citenamefont {Loewenberg}}]{cristini2003drop}%
  \BibitemOpen
  \bibfield  {author} {\bibinfo {author} {\bibfnamefont {V.}~\bibnamefont
  {Cristini}}, \bibinfo {author} {\bibfnamefont {S.}~\bibnamefont {Guido}},
  \bibinfo {author} {\bibfnamefont {A.}~\bibnamefont {Alfani}}, \bibinfo
  {author} {\bibfnamefont {J.}~\bibnamefont {B{\l}awzdziewicz}},\ and\ \bibinfo
  {author} {\bibfnamefont {M.}~\bibnamefont {Loewenberg}},\ }\bibfield  {title}
  {\enquote {\bibinfo {title} {Drop breakup and fragment size distribution in
  shear flow},}\ }\href@noop {} {\bibfield  {journal} {\bibinfo  {journal}
  {Journal of Rheology}\ }\textbf {\bibinfo {volume} {47}},\ \bibinfo {pages}
  {1283--1298} (\bibinfo {year} {2003})}\BibitemShut {NoStop}%
\bibitem [{\citenamefont {Amani}\ \emph {et~al.}(2019)\citenamefont {Amani},
  \citenamefont {Balc{\'a}zar}, \citenamefont {Castro},\ and\ \citenamefont
  {Oliva}}]{amani2019numerical}%
  \BibitemOpen
  \bibfield  {author} {\bibinfo {author} {\bibfnamefont {A.}~\bibnamefont
  {Amani}}, \bibinfo {author} {\bibfnamefont {N.}~\bibnamefont {Balc{\'a}zar}},
  \bibinfo {author} {\bibfnamefont {J.}~\bibnamefont {Castro}},\ and\ \bibinfo
  {author} {\bibfnamefont {A.}~\bibnamefont {Oliva}},\ }\bibfield  {title}
  {\enquote {\bibinfo {title} {Numerical study of droplet deformation in shear
  flow using a conservative level-set method},}\ }\href@noop {} {\bibfield
  {journal} {\bibinfo  {journal} {Chemical Engineering Science}\ }\textbf
  {\bibinfo {volume} {207}},\ \bibinfo {pages} {153--171} (\bibinfo {year}
  {2019})}\BibitemShut {NoStop}%
\bibitem [{\citenamefont {Guido}(2011)}]{guido2011shear}%
  \BibitemOpen
  \bibfield  {author} {\bibinfo {author} {\bibfnamefont {S.}~\bibnamefont
  {Guido}},\ }\bibfield  {title} {\enquote {\bibinfo {title} {Shear-induced
  droplet deformation: Effects of confined geometry and viscoelasticity},}\
  }\href@noop {} {\bibfield  {journal} {\bibinfo  {journal} {Current opinion in
  colloid \& interface science}\ }\textbf {\bibinfo {volume} {16}},\ \bibinfo
  {pages} {61--70} (\bibinfo {year} {2011})}\BibitemShut {NoStop}%
\bibitem [{\citenamefont {Ioannou}, \citenamefont {Liu},\ and\ \citenamefont
  {Zhang}(2016)}]{ioannou2016droplet}%
  \BibitemOpen
  \bibfield  {author} {\bibinfo {author} {\bibfnamefont {N.}~\bibnamefont
  {Ioannou}}, \bibinfo {author} {\bibfnamefont {H.}~\bibnamefont {Liu}},\ and\
  \bibinfo {author} {\bibfnamefont {Y.}~\bibnamefont {Zhang}},\ }\bibfield
  {title} {\enquote {\bibinfo {title} {Droplet dynamics in confinement},}\
  }\href@noop {} {\bibfield  {journal} {\bibinfo  {journal} {Journal of
  Computational Science}\ }\textbf {\bibinfo {volume} {17}},\ \bibinfo {pages}
  {463--474} (\bibinfo {year} {2016})}\BibitemShut {NoStop}%
\bibitem [{\citenamefont {Janssen}, \citenamefont {Boon},\ and\ \citenamefont
  {Agterof}(1997)}]{janssen1997influence}%
  \BibitemOpen
  \bibfield  {author} {\bibinfo {author} {\bibfnamefont {J.}~\bibnamefont
  {Janssen}}, \bibinfo {author} {\bibfnamefont {A.}~\bibnamefont {Boon}},\ and\
  \bibinfo {author} {\bibfnamefont {W.}~\bibnamefont {Agterof}},\ }\bibfield
  {title} {\enquote {\bibinfo {title} {Influence of dynamic interfacial
  properties on droplet breakup in plane hyperbolic flow},}\ }\href@noop {}
  {\bibfield  {journal} {\bibinfo  {journal} {AIChE journal}\ }\textbf
  {\bibinfo {volume} {43}},\ \bibinfo {pages} {1436--1447} (\bibinfo {year}
  {1997})}\BibitemShut {NoStop}%
\bibitem [{\citenamefont {Phillips}, \citenamefont {Graves},\ and\
  \citenamefont {Flumerfelt}(1980)}]{phillips1980experimental}%
  \BibitemOpen
  \bibfield  {author} {\bibinfo {author} {\bibfnamefont {W.~J.}\ \bibnamefont
  {Phillips}}, \bibinfo {author} {\bibfnamefont {R.~W.}\ \bibnamefont
  {Graves}},\ and\ \bibinfo {author} {\bibfnamefont {R.~W.}\ \bibnamefont
  {Flumerfelt}},\ }\bibfield  {title} {\enquote {\bibinfo {title} {Experimental
  studies of drop dynamics in shear fields: role of dynamic interfacial
  effects},}\ }\href@noop {} {\bibfield  {journal} {\bibinfo  {journal}
  {Journal of Colloid and Interface Science}\ }\textbf {\bibinfo {volume}
  {76}},\ \bibinfo {pages} {350--370} (\bibinfo {year} {1980})}\BibitemShut
  {NoStop}%
\bibitem [{\citenamefont {Stone}\ and\ \citenamefont
  {Leal}(1990)}]{stone1990effects}%
  \BibitemOpen
  \bibfield  {author} {\bibinfo {author} {\bibfnamefont {H.~A.}\ \bibnamefont
  {Stone}}\ and\ \bibinfo {author} {\bibfnamefont {L.~G.}\ \bibnamefont
  {Leal}},\ }\bibfield  {title} {\enquote {\bibinfo {title} {The effects of
  surfactants on drop deformation and breakup},}\ }\href@noop {} {\bibfield
  {journal} {\bibinfo  {journal} {Journal of Fluid Mechanics}\ }\textbf
  {\bibinfo {volume} {220}},\ \bibinfo {pages} {161--186} (\bibinfo {year}
  {1990})}\BibitemShut {NoStop}%
\bibitem [{\citenamefont {Kruijt-Stegeman}, \citenamefont {Van~de Vosse},\ and\
  \citenamefont {Meijer}(2004)}]{kruijt2004droplet}%
  \BibitemOpen
  \bibfield  {author} {\bibinfo {author} {\bibfnamefont {Y.}~\bibnamefont
  {Kruijt-Stegeman}}, \bibinfo {author} {\bibfnamefont {F.}~\bibnamefont
  {Van~de Vosse}},\ and\ \bibinfo {author} {\bibfnamefont {H.}~\bibnamefont
  {Meijer}},\ }\bibfield  {title} {\enquote {\bibinfo {title} {Droplet behavior
  in the presence of insoluble surfactants},}\ }\href@noop {} {\bibfield
  {journal} {\bibinfo  {journal} {Physics of Fluids}\ }\textbf {\bibinfo
  {volume} {16}},\ \bibinfo {pages} {2785--2796} (\bibinfo {year}
  {2004})}\BibitemShut {NoStop}%
\bibitem [{\citenamefont {Feigl}\ \emph {et~al.}(2007)\citenamefont {Feigl},
  \citenamefont {Megias-Alguacil}, \citenamefont {Fischer},\ and\ \citenamefont
  {Windhab}}]{feigl2007simulation}%
  \BibitemOpen
  \bibfield  {author} {\bibinfo {author} {\bibfnamefont {K.}~\bibnamefont
  {Feigl}}, \bibinfo {author} {\bibfnamefont {D.}~\bibnamefont
  {Megias-Alguacil}}, \bibinfo {author} {\bibfnamefont {P.}~\bibnamefont
  {Fischer}},\ and\ \bibinfo {author} {\bibfnamefont {E.~J.}\ \bibnamefont
  {Windhab}},\ }\bibfield  {title} {\enquote {\bibinfo {title} {Simulation and
  experiments of droplet deformation and orientation in simple shear flow with
  surfactants},}\ }\href@noop {} {\bibfield  {journal} {\bibinfo  {journal}
  {Chemical engineering science}\ }\textbf {\bibinfo {volume} {62}},\ \bibinfo
  {pages} {3242--3258} (\bibinfo {year} {2007})}\BibitemShut {NoStop}%
\bibitem [{\citenamefont {Milliken}\ and\ \citenamefont
  {Leal}(1994)}]{milliken1994influence}%
  \BibitemOpen
  \bibfield  {author} {\bibinfo {author} {\bibfnamefont {W.~J.}\ \bibnamefont
  {Milliken}}\ and\ \bibinfo {author} {\bibfnamefont {L.~G.}\ \bibnamefont
  {Leal}},\ }\bibfield  {title} {\enquote {\bibinfo {title} {The influence of
  surfactant on the deformation and breakup of a viscous drop: The effect of
  surfactant solubility},}\ }\href@noop {} {\bibfield  {journal} {\bibinfo
  {journal} {Journal of colloid and interface science}\ }\textbf {\bibinfo
  {volume} {166}},\ \bibinfo {pages} {275--285} (\bibinfo {year}
  {1994})}\BibitemShut {NoStop}%
\bibitem [{\citenamefont {Eggleton}\ and\ \citenamefont
  {Stebe}(1998)}]{eggleton1998adsorption}%
  \BibitemOpen
  \bibfield  {author} {\bibinfo {author} {\bibfnamefont {C.~D.}\ \bibnamefont
  {Eggleton}}\ and\ \bibinfo {author} {\bibfnamefont {K.~J.}\ \bibnamefont
  {Stebe}},\ }\bibfield  {title} {\enquote {\bibinfo {title} {An
  adsorption--desorption-controlled surfactant on a deforming droplet},}\
  }\href@noop {} {\bibfield  {journal} {\bibinfo  {journal} {Journal of colloid
  and interface science}\ }\textbf {\bibinfo {volume} {208}},\ \bibinfo {pages}
  {68--80} (\bibinfo {year} {1998})}\BibitemShut {NoStop}%
\bibitem [{\citenamefont {Roccon}, \citenamefont {Zonta},\ and\ \citenamefont
  {Soldati}(2023)}]{PhysRevFluids.8.090501}%
  \BibitemOpen
  \bibfield  {author} {\bibinfo {author} {\bibfnamefont {A.}~\bibnamefont
  {Roccon}}, \bibinfo {author} {\bibfnamefont {F.}~\bibnamefont {Zonta}},\ and\
  \bibinfo {author} {\bibfnamefont {A.}~\bibnamefont {Soldati}},\ }\bibfield
  {title} {\enquote {\bibinfo {title} {Phase-field modeling of complex
  interface dynamics in drop-laden turbulence},}\ }\href
  {https://doi.org/10.1103/PhysRevFluids.8.090501} {\bibfield  {journal}
  {\bibinfo  {journal} {Phys. Rev. Fluids}\ }\textbf {\bibinfo {volume} {8}},\
  \bibinfo {pages} {090501} (\bibinfo {year} {2023})}\BibitemShut {NoStop}%
\bibitem [{\citenamefont {Bonelli}\ \emph {et~al.}(2019)\citenamefont
  {Bonelli}, \citenamefont {Carenza}, \citenamefont {Gonnella}, \citenamefont
  {Marenduzzo}, \citenamefont {Orlandini},\ and\ \citenamefont
  {Tiribocchi}}]{Bonelli2019}%
  \BibitemOpen
  \bibfield  {author} {\bibinfo {author} {\bibfnamefont {F.}~\bibnamefont
  {Bonelli}}, \bibinfo {author} {\bibfnamefont {L.~N.}\ \bibnamefont
  {Carenza}}, \bibinfo {author} {\bibfnamefont {G.}~\bibnamefont {Gonnella}},
  \bibinfo {author} {\bibfnamefont {D.}~\bibnamefont {Marenduzzo}}, \bibinfo
  {author} {\bibfnamefont {E.}~\bibnamefont {Orlandini}},\ and\ \bibinfo
  {author} {\bibfnamefont {A.}~\bibnamefont {Tiribocchi}},\ }\bibfield  {title}
  {\enquote {\bibinfo {title} {Lamellar ordering, droplet formation and phase
  inversion in exotic active emulsions},}\ }\href
  {https://doi.org/10.1038/s41598-019-39190-6} {\bibfield  {journal} {\bibinfo
  {journal} {Scientific Reports}\ }\textbf {\bibinfo {volume} {9}},\ \bibinfo
  {pages} {2801} (\bibinfo {year} {2019})}\BibitemShut {NoStop}%
\bibitem [{\citenamefont {Negro}\ \emph {et~al.}(2019)\citenamefont {Negro},
  \citenamefont {Carenza}, \citenamefont {Lamura}, \citenamefont {Tiribocchi},\
  and\ \citenamefont {Gonnella}}]{negro2019}%
  \BibitemOpen
  \bibfield  {author} {\bibinfo {author} {\bibfnamefont {G.}~\bibnamefont
  {Negro}}, \bibinfo {author} {\bibfnamefont {L.~N.}\ \bibnamefont {Carenza}},
  \bibinfo {author} {\bibfnamefont {A.}~\bibnamefont {Lamura}}, \bibinfo
  {author} {\bibfnamefont {A.}~\bibnamefont {Tiribocchi}},\ and\ \bibinfo
  {author} {\bibfnamefont {G.}~\bibnamefont {Gonnella}},\ }\bibfield  {title}
  {\enquote {\bibinfo {title} {Rheology of active polar emulsions: from linear
  to unidirectional and inviscid flow{,} and intermittent viscosity},}\ }\href
  {https://doi.org/10.1039/C9SM01288E} {\bibfield  {journal} {\bibinfo
  {journal} {Soft Matter}\ }\textbf {\bibinfo {volume} {15}},\ \bibinfo {pages}
  {8251--8265} (\bibinfo {year} {2019})}\BibitemShut {NoStop}%
\bibitem [{\citenamefont {Cates}\ and\ \citenamefont
  {Tjhung}(2018)}]{Cates_Tjhung_2018}%
  \BibitemOpen
  \bibfield  {author} {\bibinfo {author} {\bibfnamefont {M.~E.}\ \bibnamefont
  {Cates}}\ and\ \bibinfo {author} {\bibfnamefont {E.}~\bibnamefont {Tjhung}},\
  }\bibfield  {title} {\enquote {\bibinfo {title} {Theories of binary fluid
  mixtures: from phase-separation kinetics to active emulsions},}\ }\href
  {https://doi.org/10.1017/jfm.2017.832} {\bibfield  {journal} {\bibinfo
  {journal} {Journal of Fluid Mechanics}\ }\textbf {\bibinfo {volume} {836}},\
  \bibinfo {pages} {P1} (\bibinfo {year} {2018})}\BibitemShut {NoStop}%
\bibitem [{\citenamefont {Kay}\ and\ \citenamefont
  {Kalliadasis}(2026)}]{kay2026orientablesurfactantsliquidfilms}%
  \BibitemOpen
  \bibfield  {author} {\bibinfo {author} {\bibfnamefont {T.}~\bibnamefont
  {Kay}}\ and\ \bibinfo {author} {\bibfnamefont {S.}~\bibnamefont
  {Kalliadasis}},\ }\href {https://arxiv.org/abs/2605.23789} {\enquote
  {\bibinfo {title} {Orientable surfactants on thin liquid films: A dynamic
  density-functional theory approach},}\ } (\bibinfo {year} {2026}),\ \Eprint
  {https://arxiv.org/abs/2605.23789} {arXiv:2605.23789 [cond-mat.stat-mech]}
  \BibitemShut {NoStop}%
\bibitem [{\citenamefont {Hohenberg}\ and\ \citenamefont
  {Halperin}(1977)}]{hohenberg1977theory}%
  \BibitemOpen
  \bibfield  {author} {\bibinfo {author} {\bibfnamefont {P.~C.}\ \bibnamefont
  {Hohenberg}}\ and\ \bibinfo {author} {\bibfnamefont {B.~I.}\ \bibnamefont
  {Halperin}},\ }\bibfield  {title} {\enquote {\bibinfo {title} {Theory of
  dynamic critical phenomena},}\ }\href
  {https://doi.org/10.1103/RevModPhys.49.435} {\bibfield  {journal} {\bibinfo
  {journal} {Reviews of Modern Physics}\ }\textbf {\bibinfo {volume} {49}},\
  \bibinfo {pages} {435--479} (\bibinfo {year} {1977})}\BibitemShut {NoStop}%
\bibitem [{\citenamefont {Harris}\ \emph {et~al.}(2020)\citenamefont {Harris},
  \citenamefont {Millman}, \citenamefont {Van Der~Walt}, \citenamefont
  {Gommers}, \citenamefont {Virtanen}, \citenamefont {Cournapeau},
  \citenamefont {Wieser}, \citenamefont {Taylor}, \citenamefont {Berg},
  \citenamefont {Smith} \emph {et~al.}}]{harris2020array}%
  \BibitemOpen
  \bibfield  {author} {\bibinfo {author} {\bibfnamefont {C.~R.}\ \bibnamefont
  {Harris}}, \bibinfo {author} {\bibfnamefont {K.~J.}\ \bibnamefont {Millman}},
  \bibinfo {author} {\bibfnamefont {S.~J.}\ \bibnamefont {Van Der~Walt}},
  \bibinfo {author} {\bibfnamefont {R.}~\bibnamefont {Gommers}}, \bibinfo
  {author} {\bibfnamefont {P.}~\bibnamefont {Virtanen}}, \bibinfo {author}
  {\bibfnamefont {D.}~\bibnamefont {Cournapeau}}, \bibinfo {author}
  {\bibfnamefont {E.}~\bibnamefont {Wieser}}, \bibinfo {author} {\bibfnamefont
  {J.}~\bibnamefont {Taylor}}, \bibinfo {author} {\bibfnamefont
  {S.}~\bibnamefont {Berg}}, \bibinfo {author} {\bibfnamefont {N.~J.}\
  \bibnamefont {Smith}}, \emph {et~al.},\ }\bibfield  {title} {\enquote
  {\bibinfo {title} {Array programming with numpy},}\ }\href@noop {} {\bibfield
   {journal} {\bibinfo  {journal} {nature}\ }\textbf {\bibinfo {volume}
  {585}},\ \bibinfo {pages} {357--362} (\bibinfo {year} {2020})}\BibitemShut
  {NoStop}%
\bibitem [{\citenamefont {Virtanen}\ \emph {et~al.}(2020)\citenamefont
  {Virtanen}, \citenamefont {Gommers}, \citenamefont {Oliphant}, \citenamefont
  {Haberland}, \citenamefont {Reddy}, \citenamefont {Cournapeau}, \citenamefont
  {Burovski}, \citenamefont {Peterson}, \citenamefont {Weckesser},
  \citenamefont {Bright} \emph {et~al.}}]{virtanen2020scipy}%
  \BibitemOpen
  \bibfield  {author} {\bibinfo {author} {\bibfnamefont {P.}~\bibnamefont
  {Virtanen}}, \bibinfo {author} {\bibfnamefont {R.}~\bibnamefont {Gommers}},
  \bibinfo {author} {\bibfnamefont {T.~E.}\ \bibnamefont {Oliphant}}, \bibinfo
  {author} {\bibfnamefont {M.}~\bibnamefont {Haberland}}, \bibinfo {author}
  {\bibfnamefont {T.}~\bibnamefont {Reddy}}, \bibinfo {author} {\bibfnamefont
  {D.}~\bibnamefont {Cournapeau}}, \bibinfo {author} {\bibfnamefont
  {E.}~\bibnamefont {Burovski}}, \bibinfo {author} {\bibfnamefont
  {P.}~\bibnamefont {Peterson}}, \bibinfo {author} {\bibfnamefont
  {W.}~\bibnamefont {Weckesser}}, \bibinfo {author} {\bibfnamefont
  {J.}~\bibnamefont {Bright}}, \emph {et~al.},\ }\bibfield  {title} {\enquote
  {\bibinfo {title} {Scipy 1.0: fundamental algorithms for scientific computing
  in python},}\ }\href@noop {} {\bibfield  {journal} {\bibinfo  {journal}
  {Nature methods}\ }\textbf {\bibinfo {volume} {17}},\ \bibinfo {pages}
  {261--272} (\bibinfo {year} {2020})}\BibitemShut {NoStop}%
\bibitem [{\citenamefont {Pooley}\ and\ \citenamefont
  {Furtado}(2008)}]{pooley2008eliminating}%
  \BibitemOpen
  \bibfield  {author} {\bibinfo {author} {\bibfnamefont {C.}~\bibnamefont
  {Pooley}}\ and\ \bibinfo {author} {\bibfnamefont {K.}~\bibnamefont
  {Furtado}},\ }\bibfield  {title} {\enquote {\bibinfo {title} {Eliminating
  spurious velocities in the free-energy lattice boltzmann method},}\
  }\href@noop {} {\bibfield  {journal} {\bibinfo  {journal} {Physical Review
  E—Statistical, Nonlinear, and Soft Matter Physics}\ }\textbf {\bibinfo
  {volume} {77}},\ \bibinfo {pages} {046702} (\bibinfo {year}
  {2008})}\BibitemShut {NoStop}%
\bibitem [{\citenamefont {Carenza}\ \emph {et~al.}(2019)\citenamefont
  {Carenza}, \citenamefont {Gonnella}, \citenamefont {Lamura}, \citenamefont
  {Negro},\ and\ \citenamefont {Tiribocchi}}]{Carenza2019}%
  \BibitemOpen
  \bibfield  {author} {\bibinfo {author} {\bibfnamefont {L.~N.}\ \bibnamefont
  {Carenza}}, \bibinfo {author} {\bibfnamefont {G.}~\bibnamefont {Gonnella}},
  \bibinfo {author} {\bibfnamefont {A.}~\bibnamefont {Lamura}}, \bibinfo
  {author} {\bibfnamefont {G.}~\bibnamefont {Negro}},\ and\ \bibinfo {author}
  {\bibfnamefont {A.}~\bibnamefont {Tiribocchi}},\ }\bibfield  {title}
  {\enquote {\bibinfo {title} {Lattice boltzmann methods and active fluids},}\
  }\href {https://doi.org/10.1140/epje/i2019-11843-6} {\bibfield  {journal}
  {\bibinfo  {journal} {The European Physical Journal E}\ }\textbf {\bibinfo
  {volume} {42}},\ \bibinfo {pages} {81} (\bibinfo {year} {2019})}\BibitemShut
  {NoStop}%
\bibitem [{\citenamefont {Kr{\"u}ger}\ \emph {et~al.}(2017)\citenamefont
  {Kr{\"u}ger}, \citenamefont {Kusumaatmaja}, \citenamefont {Kuzmin},
  \citenamefont {Shardt}, \citenamefont {Silva},\ and\ \citenamefont
  {Viggen}}]{kruger2017lattice}%
  \BibitemOpen
  \bibfield  {author} {\bibinfo {author} {\bibfnamefont {T.}~\bibnamefont
  {Kr{\"u}ger}}, \bibinfo {author} {\bibfnamefont {H.}~\bibnamefont
  {Kusumaatmaja}}, \bibinfo {author} {\bibfnamefont {A.}~\bibnamefont
  {Kuzmin}}, \bibinfo {author} {\bibfnamefont {O.}~\bibnamefont {Shardt}},
  \bibinfo {author} {\bibfnamefont {G.}~\bibnamefont {Silva}},\ and\ \bibinfo
  {author} {\bibfnamefont {E.~M.}\ \bibnamefont {Viggen}},\ }\href
  {https://doi.org/10.1007/978-3-319-44649-3} {\emph {\bibinfo {title} {The
  Lattice Boltzmann Method: Principles and Practice}}},\ Graduate Texts in
  Physics\ (\bibinfo  {publisher} {Springer International Publishing},\
  \bibinfo {address} {Cham},\ \bibinfo {year} {2017})\BibitemShut {NoStop}%
\bibitem [{\citenamefont {Kaoui}, \citenamefont {Harting},\ and\ \citenamefont
  {Misbah}(2011)}]{Kaoui_PRE_2011}%
  \BibitemOpen
  \bibfield  {author} {\bibinfo {author} {\bibfnamefont {B.}~\bibnamefont
  {Kaoui}}, \bibinfo {author} {\bibfnamefont {J.}~\bibnamefont {Harting}},\
  and\ \bibinfo {author} {\bibfnamefont {C.}~\bibnamefont {Misbah}},\
  }\bibfield  {title} {\enquote {\bibinfo {title} {Two-dimensional vesicle
  dynamics under shear flow: Effect of confinement},}\ }\href
  {https://doi.org/10.1103/PhysRevE.83.066319} {\bibfield  {journal} {\bibinfo
  {journal} {Phys. Rev. E}\ }\textbf {\bibinfo {volume} {83}},\ \bibinfo
  {pages} {066319} (\bibinfo {year} {2011})}\BibitemShut {NoStop}%
\bibitem [{\citenamefont {Kaoui}, \citenamefont {Krüger},\ and\ \citenamefont
  {Harting}(2012)}]{Kaoui_soft_2012}%
  \BibitemOpen
  \bibfield  {author} {\bibinfo {author} {\bibfnamefont {B.}~\bibnamefont
  {Kaoui}}, \bibinfo {author} {\bibfnamefont {T.}~\bibnamefont {Krüger}},\
  and\ \bibinfo {author} {\bibfnamefont {J.}~\bibnamefont {Harting}},\
  }\bibfield  {title} {\enquote {\bibinfo {title} {How does confinement affect
  the dynamics of viscous vesicles and red blood cells?}}\ }\href
  {https://doi.org/10.1039/C2SM26289D} {\bibfield  {journal} {\bibinfo
  {journal} {Soft Matter}\ }\textbf {\bibinfo {volume} {8}},\ \bibinfo {pages}
  {9246--9252} (\bibinfo {year} {2012})}\BibitemShut {NoStop}%
\bibitem [{\citenamefont {Meg{\'\i}as-Alguacil}, \citenamefont {Fischer},\ and\
  \citenamefont {Windhab}(2006)}]{megias2006determination}%
  \BibitemOpen
  \bibfield  {author} {\bibinfo {author} {\bibfnamefont {D.}~\bibnamefont
  {Meg{\'\i}as-Alguacil}}, \bibinfo {author} {\bibfnamefont {P.}~\bibnamefont
  {Fischer}},\ and\ \bibinfo {author} {\bibfnamefont {E.~J.}\ \bibnamefont
  {Windhab}},\ }\bibfield  {title} {\enquote {\bibinfo {title} {Determination
  of the interfacial tension of low density difference liquid--liquid systems
  containing surfactants by droplet deformation methods},}\ }\href@noop {}
  {\bibfield  {journal} {\bibinfo  {journal} {Chemical engineering science}\
  }\textbf {\bibinfo {volume} {61}},\ \bibinfo {pages} {1386--1394} (\bibinfo
  {year} {2006})}\BibitemShut {NoStop}%
\bibitem [{\citenamefont {Appleford}, \citenamefont {Sanjay},\ and\
  \citenamefont {Jalaal}(2026)}]{9dgt-rkp5}%
  \BibitemOpen
  \bibfield  {author} {\bibinfo {author} {\bibfnamefont {T.}~\bibnamefont
  {Appleford}}, \bibinfo {author} {\bibfnamefont {V.}~\bibnamefont {Sanjay}},\
  and\ \bibinfo {author} {\bibfnamefont {M.}~\bibnamefont {Jalaal}},\
  }\bibfield  {title} {\enquote {\bibinfo {title} {Rheology of two-dimensional
  dilute emulsions},}\ }\href {https://doi.org/10.1103/9dgt-rkp5} {\bibfield
  {journal} {\bibinfo  {journal} {Phys. Rev. Fluids}\ }\textbf {\bibinfo
  {volume} {11}},\ \bibinfo {pages} {033607} (\bibinfo {year}
  {2026})}\BibitemShut {NoStop}%
\bibitem [{\citenamefont {Herminghaus}\ \emph {et~al.}(2014)\citenamefont
  {Herminghaus}, \citenamefont {Maass}, \citenamefont {Krüger}, \citenamefont
  {Thutupalli}, \citenamefont {Goehring},\ and\ \citenamefont
  {Bahr}}]{C4SM00550C}%
  \BibitemOpen
  \bibfield  {author} {\bibinfo {author} {\bibfnamefont {S.}~\bibnamefont
  {Herminghaus}}, \bibinfo {author} {\bibfnamefont {C.~C.}\ \bibnamefont
  {Maass}}, \bibinfo {author} {\bibfnamefont {C.}~\bibnamefont {Krüger}},
  \bibinfo {author} {\bibfnamefont {S.}~\bibnamefont {Thutupalli}}, \bibinfo
  {author} {\bibfnamefont {L.}~\bibnamefont {Goehring}},\ and\ \bibinfo
  {author} {\bibfnamefont {C.}~\bibnamefont {Bahr}},\ }\bibfield  {title}
  {\enquote {\bibinfo {title} {Interfacial mechanisms in active emulsions},}\
  }\href {https://doi.org/10.1039/C4SM00550C} {\bibfield  {journal} {\bibinfo
  {journal} {Soft Matter}\ }\textbf {\bibinfo {volume} {10}},\ \bibinfo {pages}
  {7008--7022} (\bibinfo {year} {2014})}\BibitemShut {NoStop}%
\bibitem [{\citenamefont {Singh}, \citenamefont {Tjhung},\ and\ \citenamefont
  {Cates}(2020)}]{PhysRevResearch.2.032024}%
  \BibitemOpen
  \bibfield  {author} {\bibinfo {author} {\bibfnamefont {R.}~\bibnamefont
  {Singh}}, \bibinfo {author} {\bibfnamefont {E.}~\bibnamefont {Tjhung}},\ and\
  \bibinfo {author} {\bibfnamefont {M.~E.}\ \bibnamefont {Cates}},\ }\bibfield
  {title} {\enquote {\bibinfo {title} {Self-propulsion of active droplets
  without liquid-crystalline order},}\ }\href
  {https://doi.org/10.1103/PhysRevResearch.2.032024} {\bibfield  {journal}
  {\bibinfo  {journal} {Phys. Rev. Res.}\ }\textbf {\bibinfo {volume} {2}},\
  \bibinfo {pages} {032024(R)} (\bibinfo {year} {2020})}\BibitemShut {NoStop}%
\bibitem [{\citenamefont {Ruske}\ and\ \citenamefont
  {Yeomans}(2021)}]{PhysRevX.11.021001}%
  \BibitemOpen
  \bibfield  {author} {\bibinfo {author} {\bibfnamefont {L.~J.}\ \bibnamefont
  {Ruske}}\ and\ \bibinfo {author} {\bibfnamefont {J.~M.}\ \bibnamefont
  {Yeomans}},\ }\bibfield  {title} {\enquote {\bibinfo {title} {Morphology of
  active deformable 3d droplets},}\ }\href
  {https://doi.org/10.1103/PhysRevX.11.021001} {\bibfield  {journal} {\bibinfo
  {journal} {Phys. Rev. X}\ }\textbf {\bibinfo {volume} {11}},\ \bibinfo
  {pages} {021001} (\bibinfo {year} {2021})}\BibitemShut {NoStop}%
\end{thebibliography}%

\end{document}